\documentclass[prd,%
reprint, 
superscriptaddress,
showpacs,
nofootinbib,
 amsmath,amssymb,
 aps,
floats,
floatfix,
twocolumn]{revtex4-1}

\usepackage{graphicx}
\usepackage{dcolumn}
\usepackage{bm}
\usepackage{url}
\newcommand{\PR}[1]{\ensuremath{\left[#1\right]}} 
\newcommand{\PC}[1]{\ensuremath{\left(#1\right)}} 
\usepackage{hyperref}

\begin{document}


\title{Linear perturbations in viable $f(R)$ theories}

\author{Nelson A. Lima}
 \email{n.aguiar-lima@sussex.ac.uk}
\affiliation{Astronomy Centre, University of Sussex, Falmer, Brighton, BN1 9QH, UK}

\author{Andrew R. Liddle}%
 \email{arl@roe.ac.uk}
\affiliation{Institute for Astronomy, University of Edinburgh, Royal Observatory, Blackford Hill, Edinburgh, EH9 3HJ, UK}

\renewcommand{\abstractname}{Abstract}
\begin{abstract}
We describe the cosmological evolution predicted by three distinct $f(R)$ theories, with emphasis on the evolution of linear perturbations. The most promising observational tools for distinguishing $f(R)$ theories from $\Lambda$CDM are those intrinsically related to the growth of structure, such as weak lensing. At the linear level, the enhancement in the gravitational potential provided by the additional $f(R)$ `fifth force' can separate the theories, whereas at the background level they can be indistinguishable. Under the stringent constraints imposed on the models by Solar System tests and galaxy-formation criteria, we show that the relative difference between the models' linear evolution of the lensing potential will be extremely hard to detect even with future space-based experiments such as {\it Euclid}, with a maximum value of approximately $4 \%$ for small scales. We also show the evolution of the gravitational potentials under more relaxed local constraint conditions, where the relative difference 
between these models and $\Lambda$CDM could prove discriminating.
\end{abstract}

\pacs{98.80.-k, 95.36.+x, 04.50.Kd \hfill \today}
\maketitle


\section{\label{Int}Introduction}

Einstein's General Relativity (GR) is modern cosmology's main framework, providing a set of equations that dictate the dynamics of our Universe according to its material constituents. By them, our Universe could be expanding, static, or even collapsing. However, it is now well established that our Universe is currently undergoing an accelerated expansion which was preceded by phases of matter and radiation domination where gravitational attraction resulted in a decelerated expansion. And, at the beginning, it should have experienced a period of quasi-exponential inflation, so that any primordial spatial curvature would have been wiped out, leading to the spatially-flat and homogeneous Universe we observe.

The simplest explanation for the Universe's accelerated expansion is a cosmological constant, $\Lambda$, with a constant ratio of pressure to density (usually defined as the equation of state, $w$) equal to $-1$. Despite being in agreement with supernovae observations \cite{accel1,accel2,accel3,accel4}, data from the cosmic microwave background (CMB) \cite{cmb1,cmb2} including the recent {\it Planck} data \cite{planck1}, and large-scale structure (LSS) data \cite{lss1}, cosmologists still struggle to account for the difference between the theoretically-expected value for its energy density and the observed one. If it exists, observationally, it should account for approximately $70 \%$ of the Universe's total energy density, a value $121$ orders of magnitude smaller than that obtained from quantum field theory (for a review on $\Lambda$, see Ref.~\cite{lambdareview}). 

In light of these issues, new physics may be in order to account for that major component of our Universe, usually labelled dark energy (DE). Some theories, such as quintessence, k-essence, and so on, propose scalar fields rolling in a potential (see Ref.~\cite{quintessence} and references therein for a comprehensive review). Other theories consider higher dimensions, as in braneworld models such as the DGP model \cite{DGP,branes}, or assume that GR fails on cosmological scales and propose corrections to Einstein's action. The latter are grouped as the so-called Modified Gravity Theories (MGT), such as the Brans--Dicke scalar--tensor theory \cite{bd}, Galileon models \cite{gall}, the Fab Four \cite{fabfour}, $f(R)$ theories \cite{frreview}, and many others. For an extensive review on MGT, see Ref.~\cite{reviewall}.

In this paper, we focus on $f(R)$ models. These are modifications of the Einstein--Hilbert action through adoption of a general function $f(R)$ of the invariant Ricci scalar $R$. Even though this class of MGT might be viewed as a toy model, it is interesting as it allows for fairly general modifications of the action and appears to be one of the few that avoids the potentially-fatal Ostrogradski instability \cite{ostro}. At the very least, the study of these models can provide information on how GR may be modified and the limits to such modifications, if they prove to be necessary.

Therefore, $f(R)$ has received a great deal of attention. Since Starobinsky's first working model of inflation \cite{staro} which consisted of adding an $R^{2}$ term to the action, there have been many more attempts to develop models that are cosmologically viable explanations of the Universe's acceleration \cite{fr1,fr2,fr3,fr4}. In this work, we focus on Starobinsky's $f(R)$ model \cite{staro2}, the Hu--Sawicki model \cite{hus}, and the exponential model \cite{expo}, the reason being that these are models that provide viable cosmologies, while behaving like an effective cosmological constant in the high-curvature regime.

Due to the fourth-order nature of the equations of motion for $f(R)$, one can also apply the so-called designer approach to match any background history of the Universe \cite{fr5}. However, respecting the stringent viability conditions, which we will describe later, one is usually restricted to a cosmological evolution almost indistinguishable from the $\Lambda$CDM. Nevertheless, one can always search for modifications in the gravitational potentials by analyzing the evolution of linear perturbations in $f(R)$ theories, which has been done extensively \cite{fr5,pert1,pert2,pert3,pert4,pert41,pert42,pert5}. Due to the existence of an additional scalar degree of freedom that mediates an attractive `fifth force', there can be detectable differences on the evolution of these potentials between these theories and $\Lambda$CDM. This renders the search for modifications in the growth dynamics one of the primary goals of upcoming dark energy projects, such as the Dark Energy Survey (DES) \cite{des} and the {\it 
Euclid} mission \cite{euclid}, for instance.

In this paper, we provide, for each of the chosen $f(R)$ models, two contrasting cases of the background evolution predicted by them for a choice of parameters that either substantially violate the viability conditions or are within the observational constraints. We then study the evolution of linear perturbations for these models on different scales, noting where this diverges from $\Lambda$CDM. We note differences in the evolution of the gravitational potentials between the different models which would possibly allow them to be distinguished. We also use the designer approach to present the evolution of the linear perturbations for an $f(R)$ model with an effective equation of state equal to $\Lambda$CDM's $w = -1$ and that completely respects the viability conditions.

This article is organized as follows. In Section \ref{I}, we present the cosmological equations in the context of $f(R)$ theories and discuss the chosen models as well as the viability conditions. In Section \ref{II}, we show the linearly-perturbed equations and examine some of their generic features. In Section \ref{III}, we present the results and conclude with a brief discussion in Section \ref{conclusion}.

\section{\label{I}Cosmology in $f(R)$}

\subsection{Dynamical equations}

Our treatment of the background dynamics closely follows Refs.~\cite{pert2,frcosmo}.
The action of $f(R)$ gravity in the Jordan frame is
\begin{equation}
 S = \frac{1}{2\kappa^2} \int d^{4}x \sqrt{-g} \left[ f(R) + 2\kappa^2 \mathcal{L}_{\rm m}(\chi_i,g_{\mu \nu}) \right]\,,
\end{equation}
where $\kappa^{2} = 8 \pi G$, and $f(R)$ is a general function of the Ricci scalar, $R$, of the form $f(R) = R + \tilde{f}(R)$ , where $\tilde{f}(R)$ will be the change to GR's Einstein--Hilbert action, effectively playing the role of DE. In this frame the matter fields, $\chi_i$, will fall along the geodesics defined by the metric $g_{\mu \nu}$, since the respective Lagrangian, $\mathcal{L}$, is minimally coupled. The field equations are obtained by varying the action with respect to the metric, yielding
\begin{equation}{\label{eqmotion1}}
 f_R R_{\mu \nu} - \frac{1}{2} f g_{\mu \nu} - \PC{\nabla_\mu \nabla_\nu - g_{\mu \nu} \Box}f_R = \kappa^{2} T_{\mu \nu}\,.
\end{equation}
In this equation, $f_R \equiv \partial f / \partial R$ and $\Box \equiv g^{\mu \nu} \nabla_\mu \nabla_\nu$ is the covariant D'Alembertian. $T_{\mu \nu}$ is the energy--momentum tensor of matter, which is taken to be that of a perfect fluid
\begin{equation}{\label{conserlaw}}
 T_{\mu \nu} = \PC{\rho + p}U_\mu U_\nu + p g_{\mu \nu}\,,
\end{equation}
where $U^{\mu}$ is the fluid rest-frame four-velocity, $\rho$ is the energy density and $p$ is the pressure, related to the density by $w = p/\rho$, where $w$ is the equation of state ($w$ is $0$ for pressureless matter and $1/3$ for radiation). Due to the minimal coupling of matter to the metric, the energy--momentum tensor will obey the same conservation law as in standard GR. Adopting a flat Friedmann--Robertson--Walker (FRW) metric, $ds^2 = -dt^2 + a^{2}(t)d{\bf x}^2$, this has the well-known form
\begin{equation}
 \dot{\rho} + 3 H \PC{\rho + p} = 0\,,
\end{equation}
where overdot is differentiation with respect to time $t$ and $H = \dot{a}/a$ is the Hubble expansion factor.

The appearance of a new scalar degree of freedom in $f(R)$ theories can be seen by taking the trace of Eq.~(\ref{eqmotion1}),
\begin{equation}{\label{pot}}
 \Box{\tilde{f}_R} = \frac{1}{3}\PC{R + 2 \tilde{f} - R \tilde{f}_R} - \frac{\kappa^2}{3}\PC{\rho - 3 p} \equiv \frac{\partial V_{\textrm{eff}}}{\partial \tilde{f}_R}\,,
\end{equation}
which is a second-order differential equation for a field $\tilde{f}_R$, the scalaron \cite{staro}, with a canonical kinetic term and an effective potential $V_{\textrm{eff}}(\tilde{f}_R)$. This scalar degree of freedom can also be seen by conformally transforming the metric so that the gravitational part of the action resembles that of GR, describing the model in the Einstein frame. This renders $f(R)$ equivalent to Brans--Dicke theories with $w_{\textrm{BD}} = 0$ and a potential determined by the form of $f(R)$ where a scalar degree of freedom evolves as dark energy. Even though, in this frame, things might sometimes be conceptually simpler, the transformation also leads to a non-minimal coupling of the matter to the metric. Making explicit the dynamical equivalence between the approaches is beyond the scope of this work so, for the remainder we will stick to the Jordan frame, referring the reader to Ref.~\cite{scalar1,scalar2} and references therein for a more detailed discussion of the subject.

In order to have consistency with our knowledge from the high-redshift Universe, which is well constrained by CMB observations \cite{cmb1,cmb2}, one wants $|\tilde{f}| \ll R$ and $|\tilde{f}_R| \ll 1$ to recover standard GR with a negligible cosmological constant. In that regime, the extremum of the effective potential is located at the GR value $R = \kappa^2\PC{\rho - 3p}$. The nature of that extremum is defined by the second derivative of the potential, which can also be interpreted as the effective mass of the scalaron
\begin{equation}{\label{mass}}
 m^{2}_{\tilde{f}_R} \equiv \frac{\partial^{2} V_{\textrm{eff}}}{\partial \tilde{f}^{2}_R} = \frac{1}{3}\PR{\frac{1 + \tilde{f}_R}{\tilde{f}_{RR}} - R}\,,
\end{equation}
where $\tilde{f}_{RR}$ is the second partial derivative of $\tilde{f}$ with respect to $R$. One can then define the Compton wavelength that determines the range of the attractive fifth force mediated by the scalaron
\begin{equation}{\label{comptonr}}
 \lambda_{\rm C} = \frac{2\pi}{m_{\tilde{f}_R}}\,.
\end{equation}
One expects that on scales inside the Compton radius there is an enhancement in the gravitational potentials, which we will study later.

To obtain the background evolution relating to the different $f(R)$ models, we follow the approach taken in Ref.~\cite{frcosmo}. We start by re-writing Eq.~(\ref{eqmotion1}) as a dynamical equation for $R$, yielding
\begin{eqnarray}{\label{eqmotion2}}
  &f_R G_{\mu \nu}& - f_{RR} \nabla_\mu \nabla_\nu R - f_{RRR} \PC{\nabla_\mu R} \PC{\nabla_\nu R}  \\
 &+ g_{\mu \nu}& \PR{\frac{1}{2}\PC{R f_R - f } + f_{RR} \Box R + f_{RRR} \PC{\nabla R}^2} = \kappa^2 T_{\mu \nu}\,. \nonumber
\end{eqnarray}
Taking the trace one finds
\begin{equation}{\label{eqr}}
 \Box R = \frac{1}{3 f_{RR}}\PR{\kappa^2 T - 3 f_{RRR} \PC{\nabla R}^2 + 2f - R f_R}\,,
\end{equation}
where $T$ is the trace of the energy--momentum tensor. This can then be reinserted into Eq.~(\ref{eqmotion2}) to give
\begin{eqnarray}{\label{eqmotion3}}
 G_{\mu \nu} &=& \frac{1}{f_R}[f_{RR} \nabla_\mu \nabla_\nu R + f_{RRR} \PC{\nabla_\mu R} \PC{\nabla_\nu R}   \nonumber \\
 && - \frac{g_{\mu \nu}}{6} \PC{R f_R + f + 2\kappa^2 T} + \kappa^2 T_{\mu \nu} ]\,,
\end{eqnarray}
where $G_{\mu \nu}$ is Einstein's tensor. 

Finally, the set of equations to retrieve the cosmology for particular $f(R)$ models will come from the $t$--$t$ and $i$--$i$ Einstein's equations, as well from Eq.~(\ref{eqr}). These are
\begin{equation}{\label{eqr2}}
 \ddot{R} = -3 H \dot{R} - \frac{1}{3 f_{RR}}\PR{3 f_{RRR}\dot{R}^2 + 2f - f_R R + \kappa^2 T}\,,
\end{equation}
 for $R$, the generalization of the usual first Friedmann equation,
\begin{equation}{\label{eqmotion4}}
 H^2 + \frac{1}{f_R}\PR{f_{RR} H \dot{R} -\frac{1}{6} \PC{f_R R - f}} = - \frac{\kappa^2 T^{t}_{\;\;t}}{3 f_R}\,,
\end{equation}
and the second Friedmann equation,
\begin{equation}{\label{eqmotion5}}
 \dot{H} = -H^{2} + \frac{1}{f_R}\PC{f_{RR} H \dot{R} + \frac{f}{6} + \frac{\kappa^2 T^{t}_{\;\;t}}{3}}\,.
\end{equation}
From the last two equations, we can define an effective energy density and pressure for our $f(R)$ component behaving like DE. These correspond to Eqs.~(30) and (31) in Ref.~\cite{frcosmo}, from which one can define the $f(R)$ effective equation of state as
\begin{equation}{\label{weff1}}
 w_{\textrm{eff}} = \frac{3 H^2 - 3\kappa^2 p_{{\rm rad}} - R}{3 \PC{3 H^2 - \kappa^2 \rho}}\,,
\end{equation}
which can be written in purely geometrical terms as \cite{pert2}
\begin{equation}{\label{weff2}}
 w_{\textrm{eff}} = -\frac{1}{3} - \frac{2}{3} \frac{\PR{ H^2 \tilde{f}_R - \frac{1}{6} \tilde{f} - H \dot{\tilde{f}}_R - \frac{1}{2} \ddot{\tilde{f}}_R}}{\PR{- H^2 \tilde{f}_R - \frac{1}{6} \tilde{f} - H \dot{\tilde{f}}_R + \frac{1}{6} \tilde{f}_R R}} \,.
\end{equation}

Summing up, one can use Eqs.~(\ref{eqr2}) and (\ref{eqmotion5}) to get the cosmology for a given model starting at an initial redshift, $z_{\rm c}$, using $\Lambda$CDM as reference to obtain the initial abundances of dark matter and radiation and the value for the Hubble parameter, $H_{\rm c}$. Setting $w_{\textrm{eff}}(z_{\rm c}) \approx -1$, one can solve Eq.~(\ref{weff1}) for $R_{\rm c}$. The initial value for $\dot{R}$ is then obtained from Eq.~(\ref{eqmotion4}). This is a very good approximation as in the high-redshift Universe one expects to have $f(R) \rightarrow R - 2 \Lambda_{\textrm{eff}}^{\infty}$ in realistic $f(R)$ models. 

Due to the fourth-order nature of $f(R)$ theories, the initial value problem (or Cauchy problem) could be ill-defined, requiring one to provide initial conditions up to the third order in derivatives. However, the metric-affine $f(R)$ gravity including a matter source in the form of a perfect fluid has been shown to be equivalent to GR. Hence the initial-value problem is well formulated and well posed, as the system of equations of motion can be recast as a system of only first-order equations in time and space in the scalar field variables (see Ref.~\cite{egtreview} and references therein for a detailed discussion).

\subsection{\label{IIa}Cosmological viability of $f(R)$ models}

A great deal of work on the viability conditions of $f(R)$ theories has been done. This gives a set of restrictions that must be respected in order for $f(R)$ to have a consistent matter domination phase prior to the onset of acceleration \cite{viable2,viable3}, to meet the strict Solar System (SS) tests of gravity \cite{sstest1,sstest2}, and to provide a stable high-curvature regime where one should recover standard GR. We will only make a brief overview of these conditions.

One immediate condition, from the definition of the scalaron's effective mass, is that $\tilde{f}_{RR} > 0$ for $|R \tilde{f}_{RR}| \ll 1$, so that the scalaron in non-tachyonic. This guarantees a stable high-curvature regime with a proper matter domination phase \cite{frrlt0}. As we want to recover standard GR at early times, we need $\tilde{f} \ll R$ as $R$ increases. Together with the $\tilde{f}_{RR} > 0$ condition, one can conclude that $\tilde{f}_R<0$.

Another requirement is that $1+\tilde{f}_R > 0$. Violating this can have several consequences, such as the graviton turning into a ghost \cite{impli1}, or the Universe rapidly becoming inhomogeneous and anisotropic \cite{impli2}. A more straightforward interpretation is that this condition prevents the effective Newton's constant, rescaled from the original by $G_{\textrm{eff}} = G/(1+\tilde{f}_R)$, changing sign.

A variety of constraints have been placed on the absolute value of $\tilde{f}_R$ today, $|\tilde{f}_{R_0}|$, 
both on SS scales and Galactic scales.  Hu and Sawicki argue that Galactic structure requires it to be
smaller than about $10^{-6}$ \cite{hus}, though we note this assumes galaxy formation in $f(R)$ proceeds the same way as in GR. The tightest current observational constraints from large-scale structure and distance indicators place upper bounds on $|\tilde{f}_{R_0}|$ between $10^{-3}$ and $10^{-7}$ at the $95 \%$ confidence level \cite{cons1,cons2}. Future constraints provided by $21$cm intensity mapping are expected to place an upper limit on $|\tilde{f}_{R0}|$ around $10^{-5}$ at the same confidence level \cite{cons3}. For our purposes we adopt the conservative view that $|\tilde{f}_{R0}|$ should not exceed $10^{-4}$, in considering specific parameters within our models.

Lastly, there is the chameleon mechanism of $f(R)$ models \cite{chame1}, which is vital to pass SS tests and can also help produce a viable background expansion, as shown in Ref.~\cite{pert3}. This is deeply connected to the identification of $f(R)$ as a scalar--tensor theory, as stressed previously \cite{scalar1,scalar2}. It ensures that the additional scalar degree of freedom acquires a large mass in regions of high concentrations of matter, such as galaxies. The additional attractive fifth force is then largely suppressed. This mechanism, alongside the conditions mentioned in the previous paragraphs, should be sufficient to get a cosmologically-viable model of $f(R)$ (for a detailed discussion of fifth-force Solar System and Equivalence Principle tests in $f(R)$ gravity, see Ref.~\cite{yukreview}).

\subsection{\label{models}Realistic models of $f(R)$}

In this work, we focus on three particular $f(R)$ models that not only mimic $\Lambda$ at early times, but also at late times. The first is Starobinsky's model \cite{staro2}, which is defined by the following $f(R)$ function,
\begin{equation}
 f(R) = R + \lambda R_{S} \PR{ \PC{1 + R^{2}/R_{S}^{2}}^{-q} - 1}\,,
\end{equation}
where $R_{S} = \sigma_{\star}H_0^{2}$ is a parameter of the model that can be adjusted to fit observations or give the right cosmological evolution. We will be using $q=2$ throughout this work, and use two sets of values for the dimensionless parameters $\sigma_{\star}$ and $\lambda$ in the cases we will be considering later. Note that this model, for $q>0$, behaves like an effective cosmological constant for $R \gg R_{S}$, such that $\Lambda_{\textrm{eff}}^{\infty} \approx - \lambda R_{S}/2$.

Furthermore, to understand the cosmological evolution predicted by this and the following models, one can define, from Eq.~(\ref{eqr}), an effective potential given by  \cite{frcosmo}
\begin{equation}
V(R) = - \frac{1}{3} Rf(R) + \int^R f(x) dx \,.
\end{equation}
If one finds a solution such that $V_{R}(R_1) = 0$, then Eq.~(\ref{eqr}) will admit the constant $R_1$ value as a solution in the regime of negligible matter contribution, e.g., outside a compact object or at late times in the evolution of the Universe. According to Eq.~(\ref{eqmotion3}), $G_{\mu \nu} = -\Lambda_{\textrm{eff}}g_{\mu \nu} \equiv -g_{\mu \nu} R_1/4$, meaning that $f(R)$ theories will mimic $\Lambda$ if $R$ approaches a critical point of $V(R)$ when the energy--momentum tensor contribution is negligible, i.e.\ \mbox{$T_{\mu \nu} \approx 0$}. This corresponds to the de Sitter point where the cosmological solution is expected to asymptotically settle.

In Starobinsky's model, the potential is given by
\begin{eqnarray}{\label{vr1}}
 V(R) &=& \frac{1}{6}\PC{R^{2} - \lambda R R_{S}\frac{4 R^4 + 5 R^2 R_{S}^{2} + 3 R_{S}^{4}}{\PC{R^2 + R_{S}^{2}}^{2}}}  \nonumber \\
 && + \frac{\lambda R_{S}^{2}}{2}\arctan \frac{R}{R_{S}}\,,
\end{eqnarray}
which will be shown later when we study the background evolution predicted by the $f(R)$ models. We note that $f_{RR}$ is not positive definite, since $f_{RR} = 0$ when $R = \pm R_{S}\sqrt{2q + 1}$.

The second model is that of Hu and Sawicki \cite{hus}. The $f(R)$  function is given by
\begin{equation}
f(R) = R - m^{2}\frac{c\PC{R/m^{2}}^n}{1 + d\PC{R/m^{2}}^n}\,,
\end{equation}
where $m^{2}$, $c$, $d$ and $n>0$ are parameters of the model. Following Ref.~\cite{frcosmo}, we take $n = 4$. According to Ref.~\cite{hus}, $m^2$ is fixed from the length scales of the Universe and takes a value around $m^2 \approx 0.24 H_{0}^{2}$, which we adopt in this work. As for $c$ and $d$, these are dimensionless parameters which we fix according to Ref.~\cite{hus} so that the predicted background evolution agrees closely with $\Lambda$CDM. In one of the two sets of parameters considered for this model, we also fix them so that $\tilde{f}_{R_{0}}$ is close to the viable range.

Again, we note that this model effectively behaves like a cosmological constant in the early high curvature Universe, such that $f(R) = R - 2 \Lambda_{\textrm{eff}}^{\infty} \equiv R - (1/2)(m^{2}c/d)$. The corresponding potential, however, does not have a reasonable analytic form. We will show it later, and it will become evident that the Ricci scalar is able to settle into a minimum at the late-time evolution. Also, even though $f_{R}$ and $f_{RR}$ are not positive definite, we do not face that situation in the obtained expansion histories.

The last model analyzed is the $f(R)$ exponential model \cite{expo}. It is defined by
\begin{equation}
 f(R) = R + \lambda R_{\star} \PC{e^{-R/R_\star} - 1}\,, 
\end{equation}
where $\lambda$ is a dimensionless parameter and $R_{\star}$ is a characteristic scale of the model, playing a similar role as $m^2$ and $R_{S}$ in the previous models. Like the previous models, this has the property of developing  an effective cosmological constant at early times, such that $\Lambda_{\textrm{eff}}^{\infty} = \lambda R_{\star}/2$. For $\lambda >0$, $f_{RR}$ is positive definite, while $1 + \tilde{f}_{R}$ will be positive as long $R>R_{\star} \ln \lambda$, which is assured in the background evolutions obtained in this work. The potential, $V(R)$, is
\begin{equation}
 V(R) = \frac{R_{\star}^{2}}{6} \PR{\tilde{R}\PC{\tilde{R} - 4 \lambda} - 2 \lambda \PC{\tilde{R} + 3}e^{-\tilde{R}}}\,,
\end{equation}
where $\tilde{R} \equiv R/R_{\star}$. Plots of this will be shown later but, as stated in Ref.~\cite{expo}, for $\lambda>0$, this potential has a local maximum at $R=0$ and a global minimum at \mbox{$R_1 >0$} needed to have a non-vanishing cosmological constant where our solution settles asymptotically in future time.

\section{\label{II}Linear perturbations in $f(R)$}

The evolution of linear perturbations in $f(R)$ models has been derived and thoroughly analyzed in Refs.~\cite{pert1,pert2,pert3,pert4,pert5}. Here, it will suffice to present the reader with the equations that are useful for this work and briefly overview their possible implications.

Working in the Jordan frame, we will be considering scalar perturbations of the metric given by the standard form
\begin{equation}{\label{permetric}}
 ds^2 = -a^{2}(\tau)\PR{ \PC{1 + 2\Psi}d\tau^{2} + \PC{1 - 2\Phi}d{\bf x}^{2}}\,,
\end{equation}
where $\tau$ is the conformal time, related to the coordinate time by $dt = a d\tau$; $\Psi$ and $\Phi$ are small scalar perturbations of the FRW metric that are both time and space dependent. As for the energy--momentum tensor, we consider the usual first-order expansion given by
\begin{eqnarray}{\label{perttensor}}
T^{0}_{\;\;0} &=& - \rho\PC{1+\delta} \,; \nonumber \\
T^{0}_{\;\;i} &=& -\PC{\rho + p}v_i  \,; \\
T^{i}_{\;\;j} &=& \PC{p + \delta p}\delta^{i}_{j} + \pi^{i}_{\;j}\,, \nonumber
\end{eqnarray}
where $\delta \equiv \delta \rho/\rho$ is the density contrast, $v_i$ is the velocity field, $\delta p$ is the pressure perturbation, and $\pi^{i}_{\;j}$ is the traceless part of the energy--momentum tensor. The perturbed energy--momentum conservation equations, since matter is minimally coupled in the $f(R)$ Lagrangian, have the same form as in standard GR:
\begin{equation}{\label{conservation1}}
\delta^{\prime} + \frac{k}{aH}V - 3\PC{1+w} \Phi^{\prime} + 3\PC{\frac{\delta p}{\delta \rho} -w}\delta = 0\,, 
\end{equation}
for the $t$--$t$ component, and
\begin{equation}{\label{conservation2}}
 V^{\prime} + \PC{1-3w}V - \frac{k}{aH}\PC{\frac{\delta p}{\delta \rho} - \frac{\Pi}{\delta}}\delta - \frac{k}{aH}\PC{1+w}\Psi = 0\,,
\end{equation}
for the individual matter components. $V$ is the scalar velocity potential, whose gradient gives $v_{i}$, and $\Pi$ is the scalar part of the anisotropic stress defined by $\rho \Pi \equiv (\hat{k}^{i}\hat{k}_{j} - 1/3\hspace{0.3 mm}\delta^{i}_{\;j})\pi^{i}_{\;j}$. Primes denote derivatives with respect to $\log a$, and $k$ is the comoving wavenumber of the expansion of the perturbed quantities in Fourier space, where the different modes evolve independently.

The full set of linearly-perturbed equations for $f(R)$ can be seen in Refs.~\cite{pert1,pert2}. The anisotropy, or space-off diagonal equation is given by \cite{pert1,pert2}
\begin{equation}{\label{lpert1}}
\Phi - \Psi = \frac{9}{2}\frac{a^{2}}{k^{2}}E_{n} \Pi_{n} - \tilde{f}_{R}\PC{\Phi - \Psi} + \tilde{f}_{RR} \delta R \,,
\end{equation}
where $\delta R$ is the linear perturbation of the Ricci scalar and $E_n$ is the density of the $n$-th matter component as a fraction of the {\em present-day} critical density. The repeated indices denote a sum over the matter fields. Neglecting any anisotropic contribution from matter fields, hence setting $\Pi_{n} = 0$, one gets the following relation between the gravitational potentials
\begin{equation}{\label{lpert2}}
 f_R\PC{\Phi - \Psi} = \tilde{f}_{RR} \delta R \,,
\end{equation}
where $f_R \equiv 1 + \tilde{f}_R$. This equation already presents a dynamical departure from standard GR, where the anisotropy equation just yields the constraint $\Psi = \Phi$. Note that this limit is recovered when \mbox{$\tilde{f} = 0$}, as expected.

The Poisson equation is given by \cite{pert1,pert2}
\begin{eqnarray}{\label{poisson}}
  f_R\frac{k^2}{a^2}\Phi = &-&\frac{3}{2} E_{n} \Delta_{n} + \frac{1}{2}\frac{k^2}{a^2}\tilde{f}_{RR}\delta R \nonumber \\
  &-&\frac{3}{2}H^{2} \tilde{f}_{R}^{\prime} \PC{\Psi + \Phi^{\prime}} -\frac{3}{2} H H^{\prime} \tilde{f}_{RR}\delta R \,,
\end{eqnarray}
where it becomes clear that the presence of the modified gravity term in the action adds extra dynamical terms to the evolution equations of the Newtonian potentials. In standard GR, this would just be an algebraic relation between $\Psi$ and the comoving matter density perturbation $\Delta_{n}$. The latter is defined as
\begin{equation}{\label{comovden}}
 \Delta_{n} \equiv \delta_{n} + 3\frac{aH}{k}(1+w_{n})V_{n} \,.
\end{equation}

Following Ref.~\cite{pert2}, one may choose instead to evolve the following variables:
\begin{equation}
 \chi = \tilde{f}_{RR} \delta R, \quad \Phi_+ = \frac{\Phi + \Psi}{2} \,,
\end{equation}
where $\chi$ is the slip between the Newtonian potentials and $\Phi_+$ is the lensing potential that is responsible for such effects as the Sachs--Wolfe effect in the CMB and weak lensing of distant galaxies. Hence, Eq.~(\ref{lpert2}) becomes a simple constraint equation, and any $\chi \neq 0$ will indicate a departure from standard GR. The evolution equations for $\Phi$ and $\Psi$ will then be obtained from the Poisson equation and from the perturbed $i$--$0$ Einstein equation (or the momentum equation). Neglecting any contribution from the radiation component, these are given by \cite{pert2}
\begin{equation}{\label{pertevol1}}
\Phi_{+}^{\prime} = \frac{3}{2}\frac{E_{\rm m} V_{\rm m}}{Hkf_{R}} - \PC{1 + \frac{1}{2}\frac{f_{R}^{\prime}}{f_R}}\Phi_{+} + \frac{3}{4}\frac{f_{R}^{\prime}}{f_{R}^{2}}\chi \,,
\end{equation}
and
\begin{eqnarray}{\label{pertevol2}}
 \chi^{\prime} = &-&\frac{2 E_{\rm m} \Delta_{\rm m}}{H^{2}} \frac{f_R}{f_{R}^{\prime}} + \PC{1 + \frac{f_{R}^{\prime}}{f_R} - 2\frac{H^{\prime}}{H}\frac{f_R}{f_{R}^{\prime}}}\chi  \nonumber \\
&-& 2f_R\Phi_{+}^{\prime} -2f_{R}\PC{1 + \frac{2}{3}\frac{k^2}{a^2 H^2} \frac{f_R}{f_{R}^{\prime}}} \,,
\end{eqnarray}
where the subscript `m' stands for ordinary matter.

At early times, the effect of $\tilde{f}(R)$ on the overall background evolution of the Universe is negligible. Therefore, for modes inside the horizon ($k > aH$) but way outside the Compton radius, $\lambda_{\rm C}$ (which is suppressed at this point since $\tilde{f}_{RR} \rightarrow 0$), one expects the evolution of the gravitational potentials to exhibit the same behavior as they do in standard GR. Hence, the lensing potential is expected to remain constant. Then, as $\lambda_{\rm C}$ increases, the Fourier mode eventually enters the radius defined by it and one should observe an enhancement in the perturbed potentials due to the attractive fifth force. Finally, at late times in the cosmological evolution, given the background expansion due to the presence of an effective cosmological constant, the Newtonian potentials should decay, as in GR.

Due to the oscillations of the linear perturbation of the Ricci scalar, $\delta R$, in $f(R)$ models \cite{staro2}, which can have catastrophic consequences relating to particle production, $\chi$ will oscillate as well with an amplitude and frequency proportional to the squared mass of the scalaron, $m_{\tilde{f}_R}^{2}$ \cite{pert5}. Nonetheless, as will be seen in the next section, these oscillations are quite suppressed due to the very small values of $\tilde{f}_{RR}$, and their effect on the evolution of the gravitational potentials is practically negligible.

\section{\label{III}Results}

Here we present the evolution of the background history and of the linear perturbations predicted by the three distinct $f(R)$ theories considered in this work. For each model, we will consider two cases:
\begin{enumerate}
 \item We choose parameters to result in a $|\tilde{f}_{R_0}| \approx 1 \times 10^{-4}$, within the observational constraints. We will use a numeral subscript~$1$ when referring to quantities obtained with this set of parameters.
 \item The value obtained for $|\tilde{f}_{R_0}|$ is in disagreement with the corresponding theoretical and observational bounds, even though the predicted background evolution is very close to $\Lambda$CDM. The numeral subscript $2$ will refer to this case.
\end{enumerate}
Additionally, we present results for an $f(R)$ model with an effective equation of state equal to $-1$ throughout the cosmological evolution and $|\tilde{f}_{R0}| \approx 10^{-6}$ using the designer 
approach mentioned before and detailed in Ref.~\cite{fr5}.

To obtain the background evolution predicted by each model, we have set the initial conditions at $z_{\rm c} = 10$, using a present-day value of $\Omega_{{\rm m}0} = 0.27$ and assuming a flat cosmology. We have ensured the numerical present-day value obtained for the Hubble parameter was in better than $1 \%$ agreement with the input $H_{0} = 72$ km s$^{-1}$Mpc$^{-1}$. Reaching higher redshifts with exact integration is not possible due to the rapid oscillations in $w_{\rm eff}$ around the phantom divide $w_{\rm eff} = -1$, which complicates the numerical treatment \cite{wosc1,wosc2}. In order to start the evolution of perturbations at high redshift, we therefore assumed that between $z_{\rm i} = 1000$ and $z_{\rm c} = 10$ the equation of state can be replaced with its time-averaged value of $-1$. Then, from $z_{\rm c} = 10$ to the present time, we use the form for $w_{\rm eff}$ recovered from the full background evolution.

For the evolution of the linear perturbations, the initial conditions were defined as in Ref.~\cite{pert2}, assuming again a flat cosmology with $\Omega_{{\rm m}0} = 0.27$. We started the evolution from $z_{\rm i} = 1000$ and the initial values of $\Phi_{+}$ and $\chi$ were $-1$ and $0$, respectively. Since the deviations from standard GR are small at this epoch, the initial conditions for $v_{\rm m}$ and $\Delta_{\rm m}$ are
\begin{equation}{\label{inicond}}
 v_{{\rm m},{\rm i}} = \frac{2k}{3aH}\Phi_{+}\quad ; \quad \Delta_{{\rm m},{\rm i}} = -\frac{2k^{2}}{3a^{2}H^{2}}\Phi_{+} \,.
\end{equation}

\subsection{\label{model1}Starobinsky model}

\begin{figure}[t!]
\begin{center}$
\begin{array}{c}
\includegraphics[scale = 0.40]{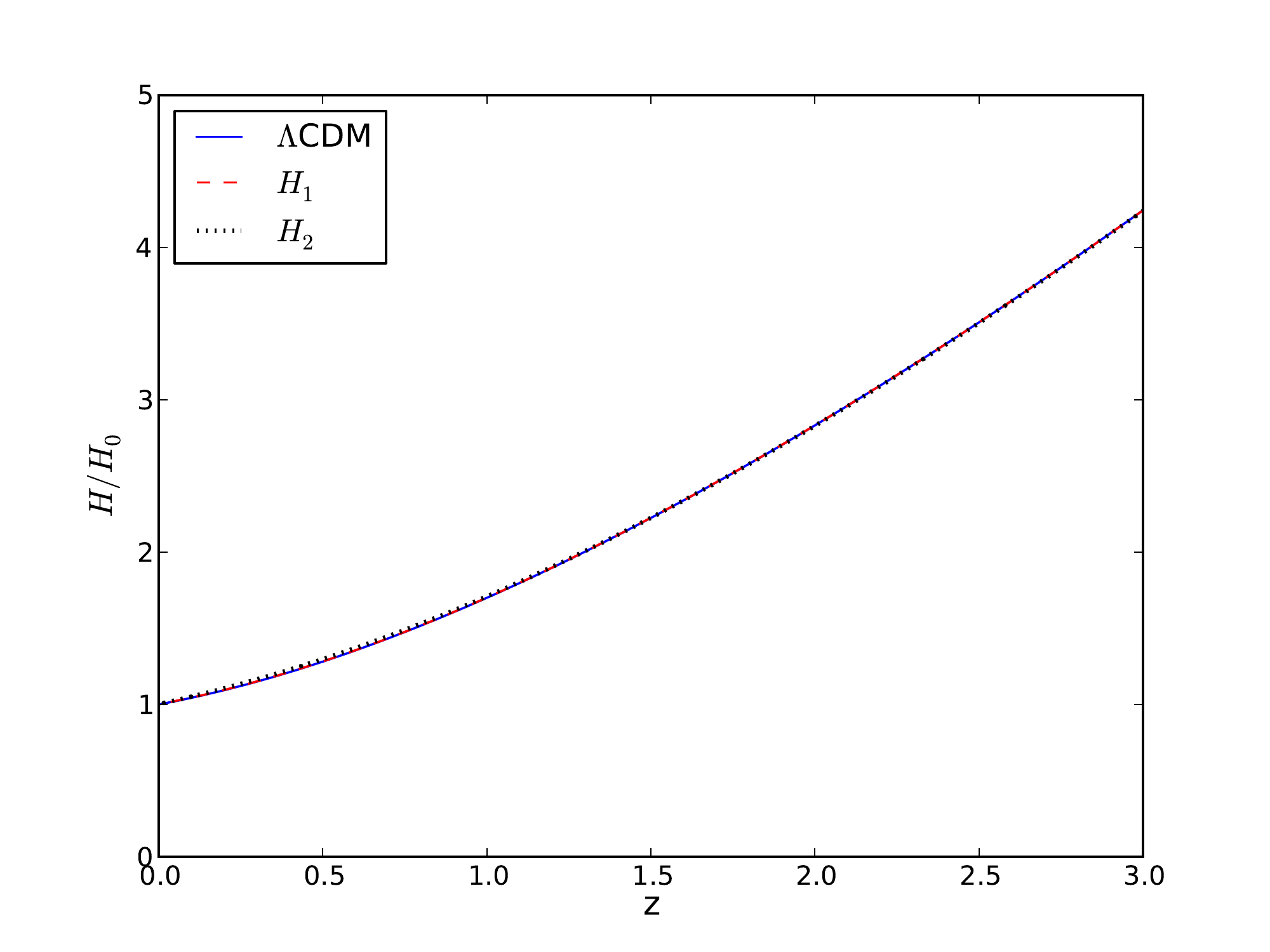} \\
\includegraphics[scale = 0.40]{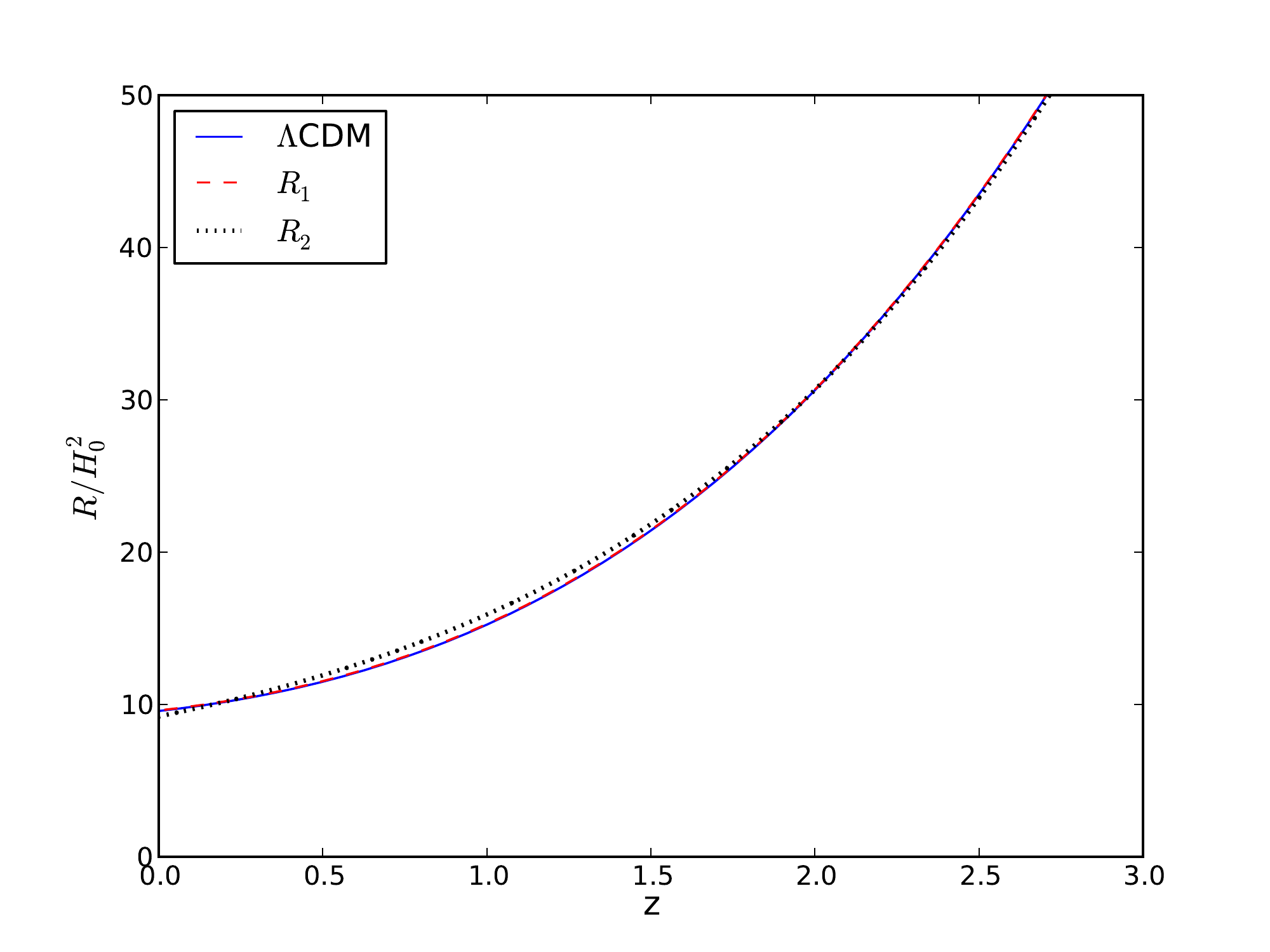} \\
\includegraphics[scale = 0.40]{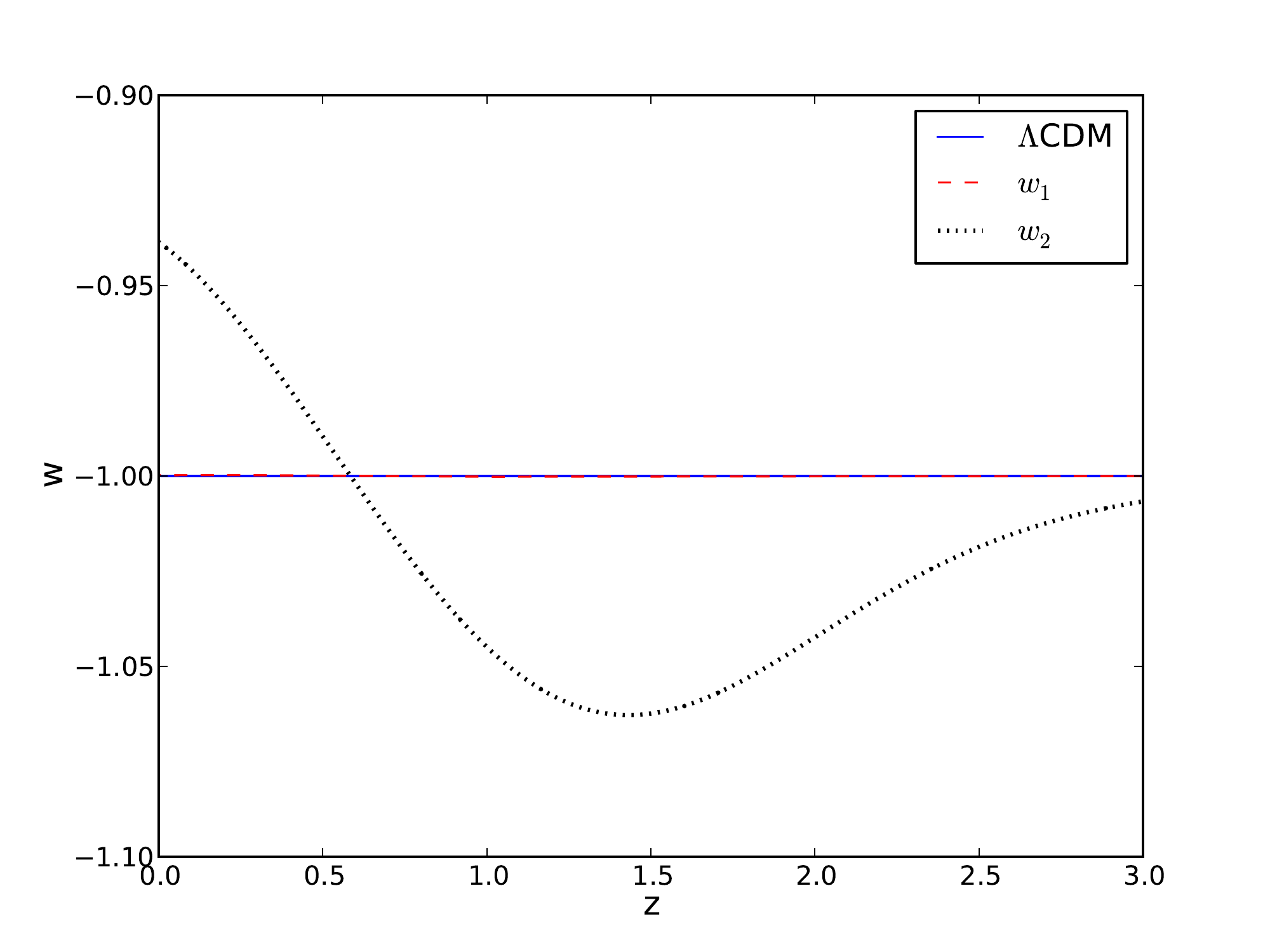}
\end{array}$
\end{center}
\caption{\label{backstaro}The Hubble parameter, Ricci scalar, and $f(R)$ effective equation of state as a function of redshift, $z$, for the Starobinsky model, compared to the $\Lambda$CDM model.}
\end{figure}

For this model, we have chosen $R_{S1} = 0.83$ and $\lambda_1 = 5.3$, and $R_{S2} = 4.17$ and $\lambda_2 = 1.0$. The latter values were used in Ref.~\cite{frcosmo}. Using the formalism described in Section \ref{I}, we start by presenting, in Fig.~\ref{backstaro}, the evolution of the background Hubble expansion factor and Ricci scalar as a function of redshift $z$. We also plot the evolution of this model's effective equation of state, as defined by Eqs.~(\ref{weff1}) or (\ref{weff2}).

In Fig.~\ref{backstaro}, we can see a typical feature of viable $f(R)$ models that satisfy $\tilde{f}_{RR} >0$, which is the phantom crossing in the equation of state. This has been emphasized in several previous works \cite{frcosmo,hus,frphantom}, and is more easily seen in $w_2$. This also happens with $w_1$, though with a very much smaller amplitude, given the fact that this equation of state is practically indistinguishable from $\Lambda$CDM's.

\begin{figure}[t!]
 $\includegraphics[scale = 0.40]{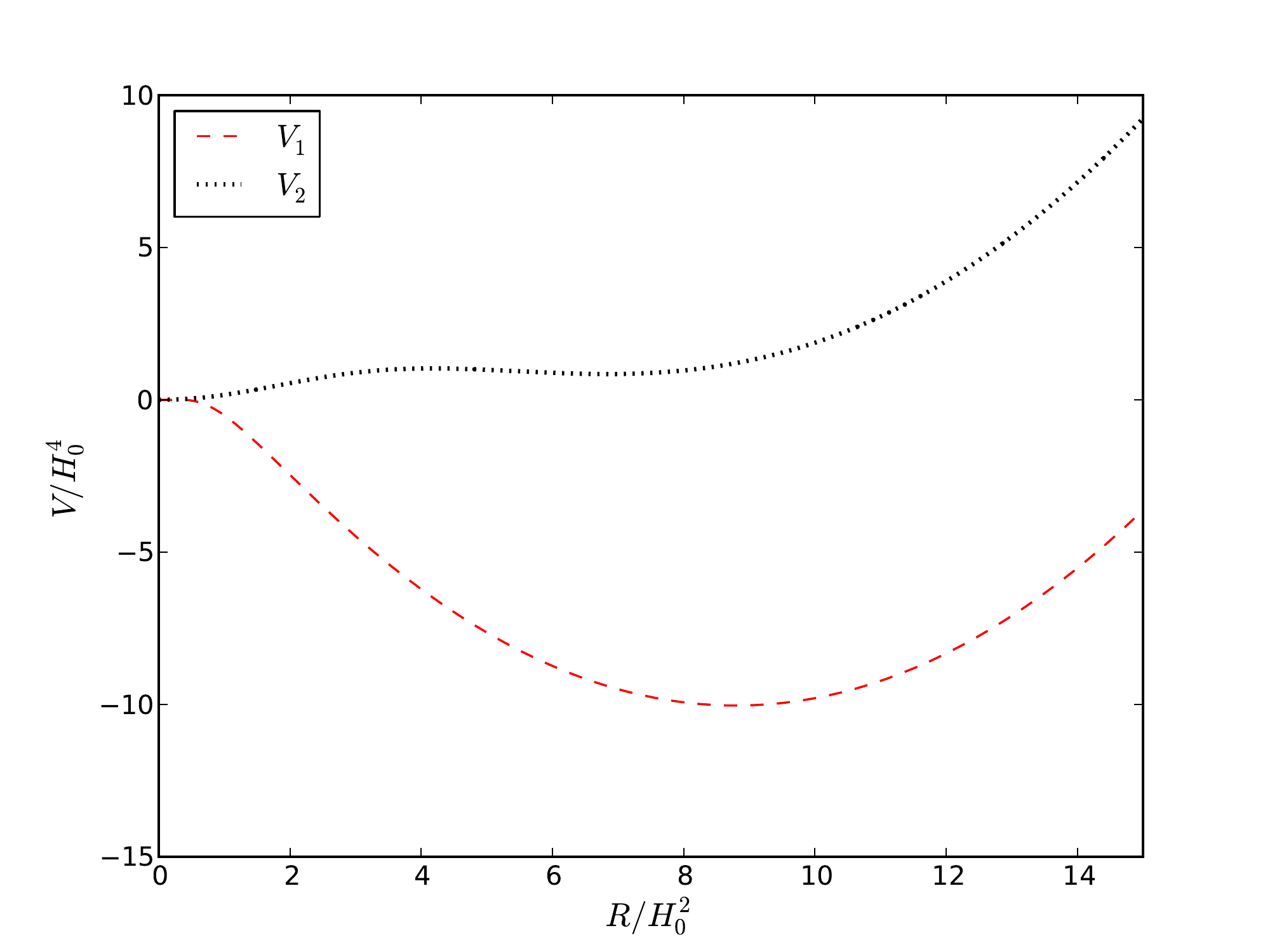}$
 \caption{\label{pot_staro} The $V(R)$ potential as a function of the Ricci scalar for the Starobinsky model.}
\end{figure}

It is interesting that two different sets of parameters yield such similar background histories, and that these are close to the cosmological evolution predicted by $\Lambda$CDM. This is expected, since the model is designed to yield a negligible cosmological constant in the high-redshift Universe and settle in a stable de Sitter point in the future. The sets yield present-day values of $H$ and $R$ that are very close to each other. However, as one can see in Fig.~\ref{pot_staro}, they should disagree in the distant future, as $R_{1_0}$ is already close to the de Sitter minimum of the $V(R)$ potential, where it will settle, while for the second set of parameters the solution is still moving towards the respective minimum, at a smaller value of $R$.

\begin{figure}[t!]
\begin{center}$
\begin{array}{c}
\includegraphics[scale = 0.40]{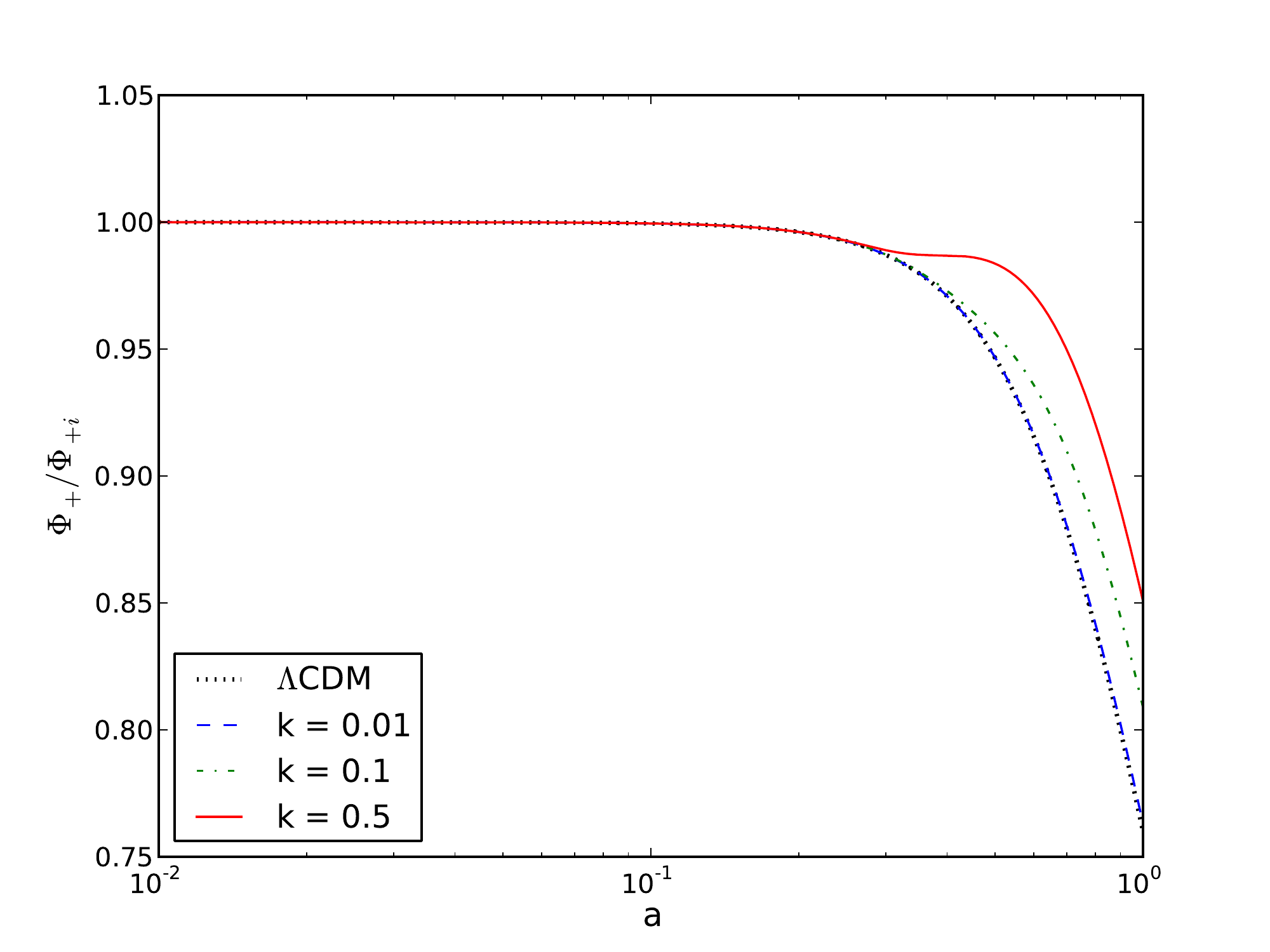} \\
\includegraphics[scale = 0.40]{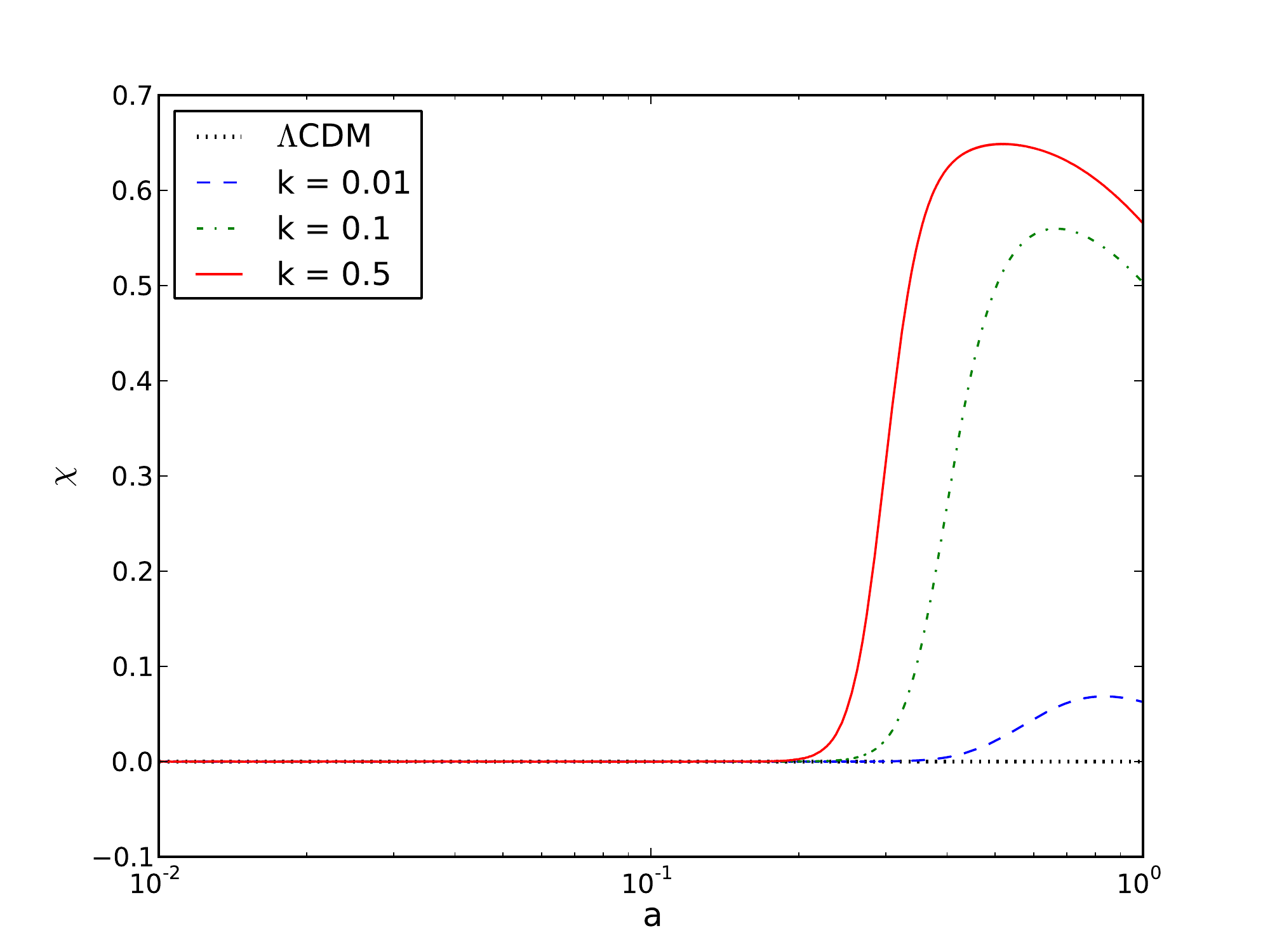}
\end{array}$
\end{center}
\caption{\label{pertstaro1} The lensing potential $\Phi_{+}$ and $\chi$ as a function of the scale factor, $a$, for the Starobinsky model, for the first set of parameters.}
\end{figure}

Despite the subtle differences in the background evolution, note that the values obtained for $\Lambda_{\textrm{eff}}$ at the de Sitter limit and in the high-curvature regime are very close to each other for both sets of parameters. Hence, for the first set, we have $\Lambda_{\textrm{eff}}^{\textrm{de\, Sitter}} \approx 2.1 H_{0}^{2}$ and $\Lambda_{\textrm{eff}}^{\infty} \approx 2.2 H_{0}^{2}$, while for the second set we have $\Lambda_{\textrm{eff}}^{\textrm{de\, Sitter}} \approx 1.7 H_{0}^{2}$ and $\Lambda_{\textrm{eff}}^{\infty} \approx 2.1 H_{0}^{2}$ \cite{frcosmo}.

Throughout the evolution of this model, we have not reached a singular point where $f_{RR}$ reached zero and then changed sign, hence the stability of the solutions is guaranteed. Also, we have obtained $\tilde{f}_{R0} \approx -1 \times 10^{-4}$ for the first set of parameters, and $\tilde{f}_{R0} \approx -4\times10^{-2}$ for the second set of parameters, in agreement with Ref.~\cite{frcosmo}.

\begin{figure}
\begin{center}$
\begin{array}{c}
\includegraphics[scale = 0.40]{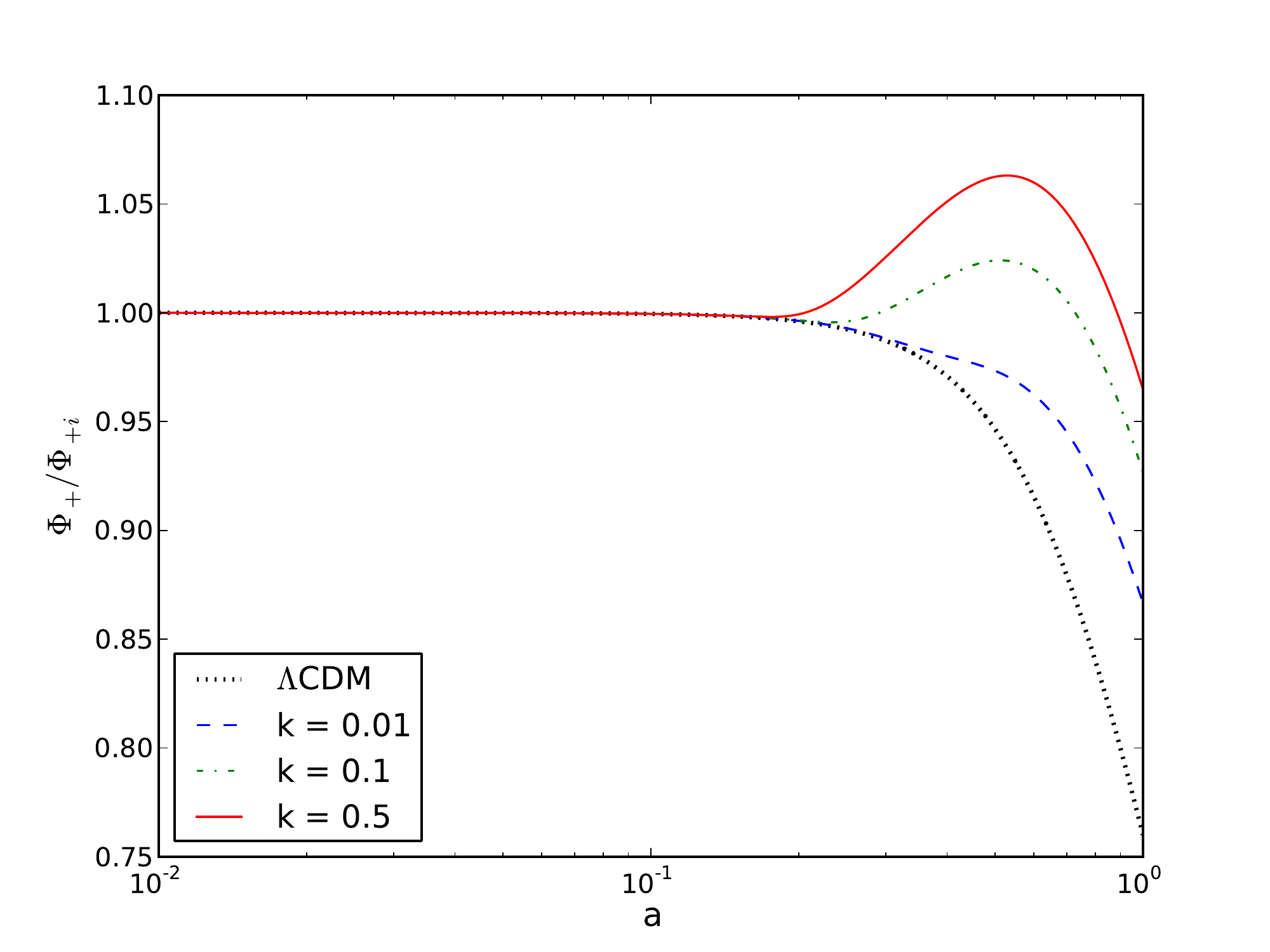} \\
\includegraphics[scale = 0.40]{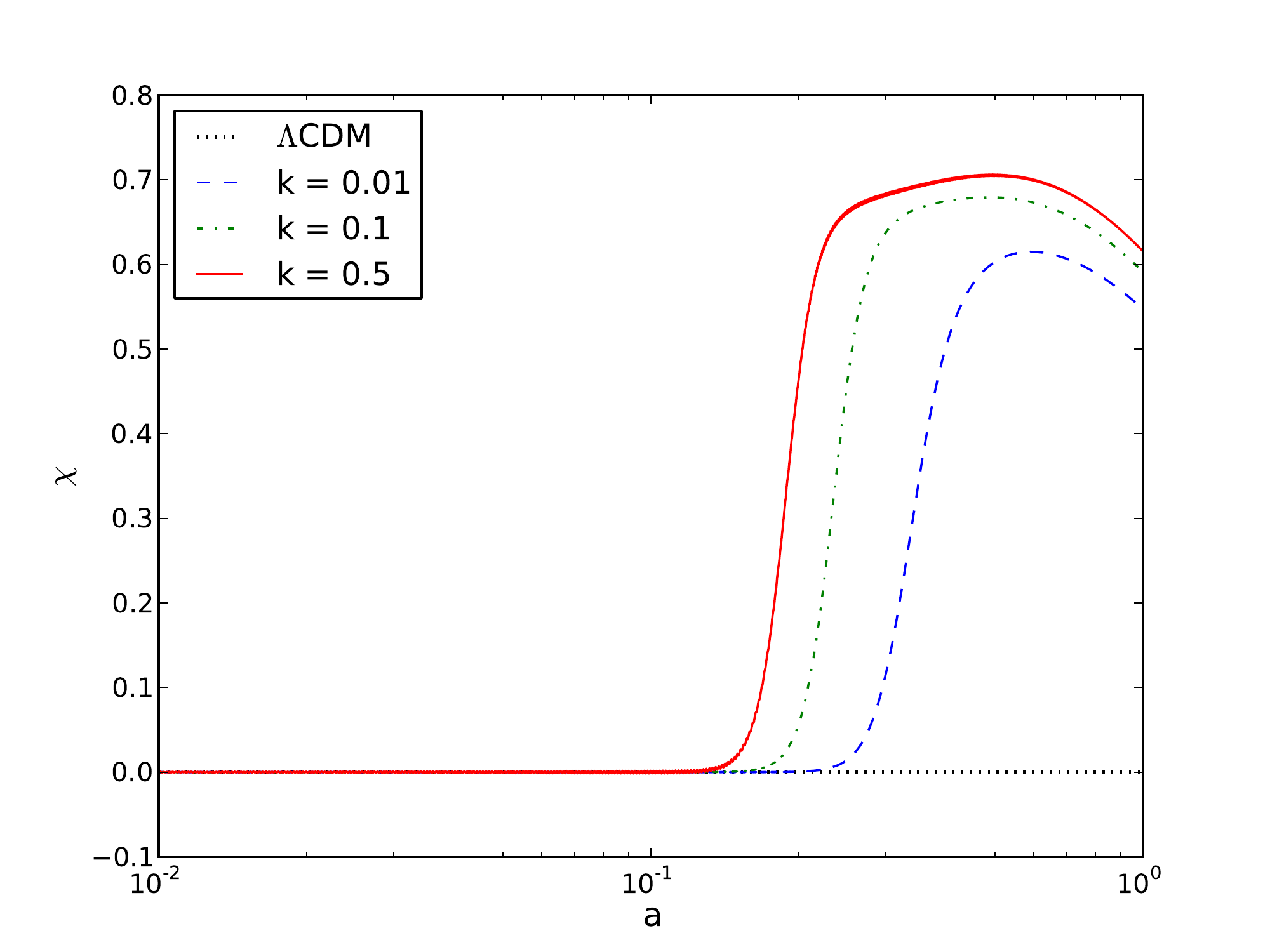}
\end{array}$
\end{center}
\caption{\label{pertstaro2}The lensing potential $\Phi_{+}$ and $\chi$ as a function of $a$ for the Starobinsky model, for the second set of parameters.}
\end{figure}

Figures \ref{pertstaro1} and \ref{pertstaro2} show the evolution of the linear perturbations in this model. In contrast to the $\Lambda$CDM case where the evolution is independent of scale, the potentials evolve differently depending on length scale. For both cases considered, the enhancement of the perturbations is stronger at smaller scales, i.e., higher wavenumber $k$.\footnote{$k$ is in units of $h/\textrm{Mpc}$, where $h$ is $H_{0}/100$ and $H_{0}$ is the present-day value of the Hubble parameter. In this work, we have taken $H_{0} = 72 \, {\rm km} \, {\rm s}^{-1} {\rm Mpc}^{-1}$.} This is because the modes corresponding to smaller scales enter  the range of action of the fifth force, defined by the Compton wavelength in Eq.~(\ref{comptonr}), sooner, the latter being dependent on the scalaron's mass defined by Eq.~(\ref{mass}). 

\begin{figure}
 $\includegraphics[scale = 0.40]{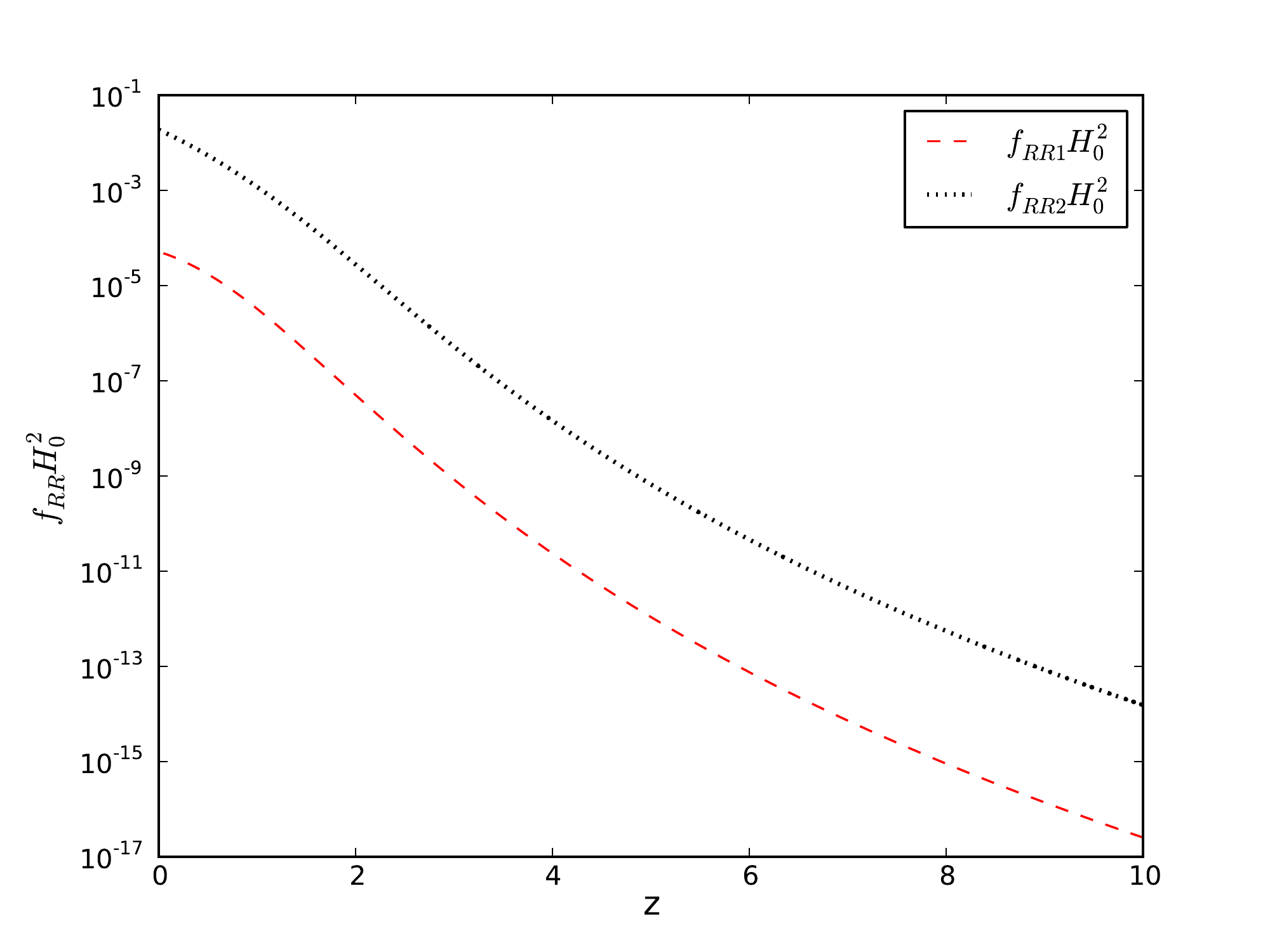}$
 \caption{\label{frr_staro}The form of $f_{RR}$ as a function of redshift for the two cases considered in the Starobinsky model.}
\end{figure}

Hence, in the high-curvature/redshift regime, the fifth force is quite suppressed as the mass is very large, since $f_{RR} \equiv \tilde{f}_{RR} \rightarrow 0$, as one can observe in Fig.~\ref{frr_staro}. Therefore, the evolution of the perturbed gravitational potentials is similar to that in scale-invariant $\Lambda$CDM with standard GR. The exception is the high-frequency oscillations in $\chi$ which depend on the mass of the scalaron, and which cannot be resolved by eye
 
Later in the evolution, when $f_{RR}$ starts to rise, the mass of the scalaron decreases, the Compton wavelength increases, and the modes start to enter it. Accordingly the evolution of the linear perturbations starts to deviate from standard GR. Inevitably, the enhancement in the perturbed potentials is suppressed by the background accelerated expansion when $z$ approaches zero.

Lastly, we note that the difference in $\tilde{f}_{R0}$ between the sets of parameters translates into significant differences in the evolution of the perturbations. The evolution of the respective $f_{RR}$ is crucial for understanding this. Looking at Fig.~\ref{frr_staro}, one sees that the sooner $f_{RR}$ starts to increase, the greater is the enhancement in the perturbed potentials. Hence, the evolution of the linear perturbations for the second set of potentials has a greater enhancement, translated to an actual growth in $\Phi_{+}$. For the first set of parameters, not only does the enhancement kick in later, but the magnitude of $f_{RR}$ remains very small throughout the evolution. Hence, for instance, $\Phi_{+}$ will not necessarily grow, as the effect of the fifth force will suffice only to resist the expanding background. In either of the cases, nevertheless, the differences from the scale-invariant $\Lambda$CDM are noticeable by eye, particularly for the smaller scales.

\begin{figure}[t!]
\begin{center}$
\begin{array}{c}
\includegraphics[scale = 0.40]{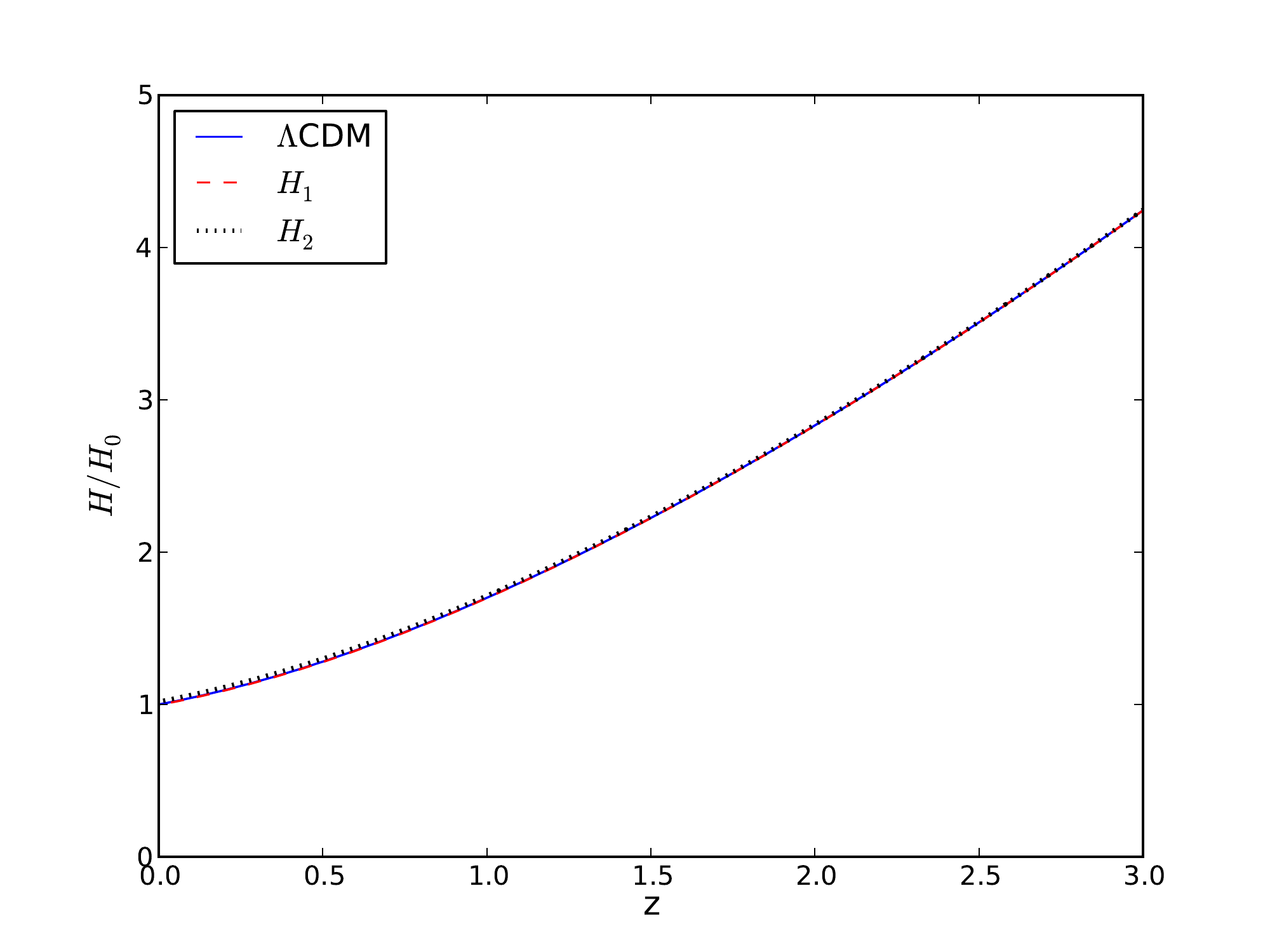} \\
\includegraphics[scale = 0.40]{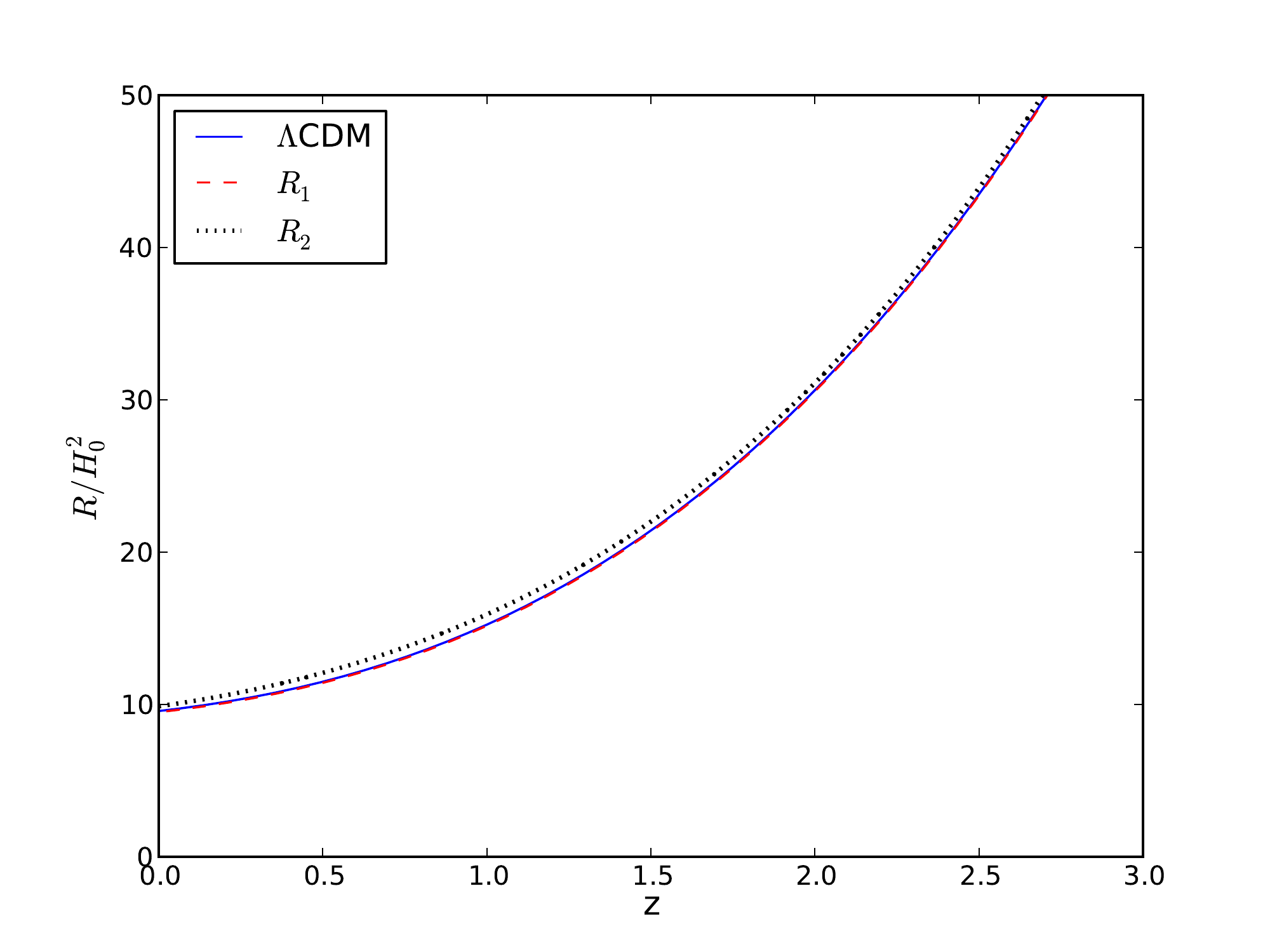} \\
\includegraphics[scale = 0.40]{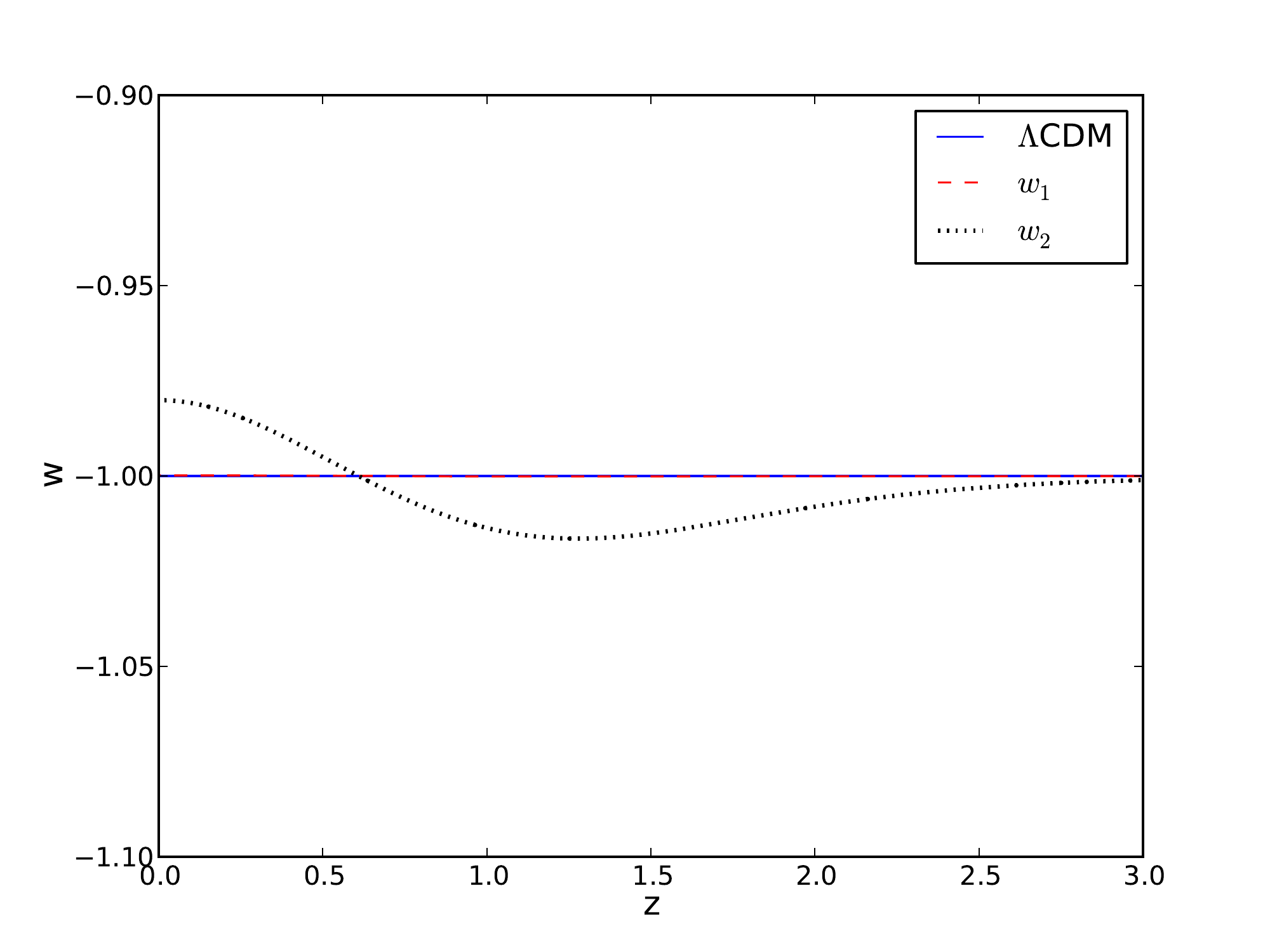}
\end{array}$
\end{center}
\caption{\label{backhu} The Hubble parameter, Ricci scalar, and $f(R)$ effective equation of state for the Hu--Sawicki model, compared to the $\Lambda$CDM model.}
\end{figure}

\subsection{\label{model2}Hu--Sawicki model}

We start with Fig.~\ref{backhu}, which shows the evolution of the Hubble expansion factor, the Ricci scalar and the effective equation of state of $f(R)$ as a function of $z$. We have used $c_{1} = 0.190$, $d_{1} = 0.0105$, $c_{2} = 1.25 \times 10^{-3}$ and $d_{2} = 6.56 \times 10^{-5}$, the latter defined as in Ref.~\cite{frcosmo}. As in the previous model, one can observe that the distinct sets of parameters yield two very similar background histories. Both the Hubble parameter, $H$, and the Ricci scalar, $R$, present an evolution as a function of redshift that is very close to that predicted by $\Lambda$CDM. The main difference lies in the evolution of the effective equation of state, where we have the usual phantom crossing, which is particularly noticeable for the second set of parameters and negligible for the first set.

\begin{figure}[t!]
 $\includegraphics[scale = 0.40]{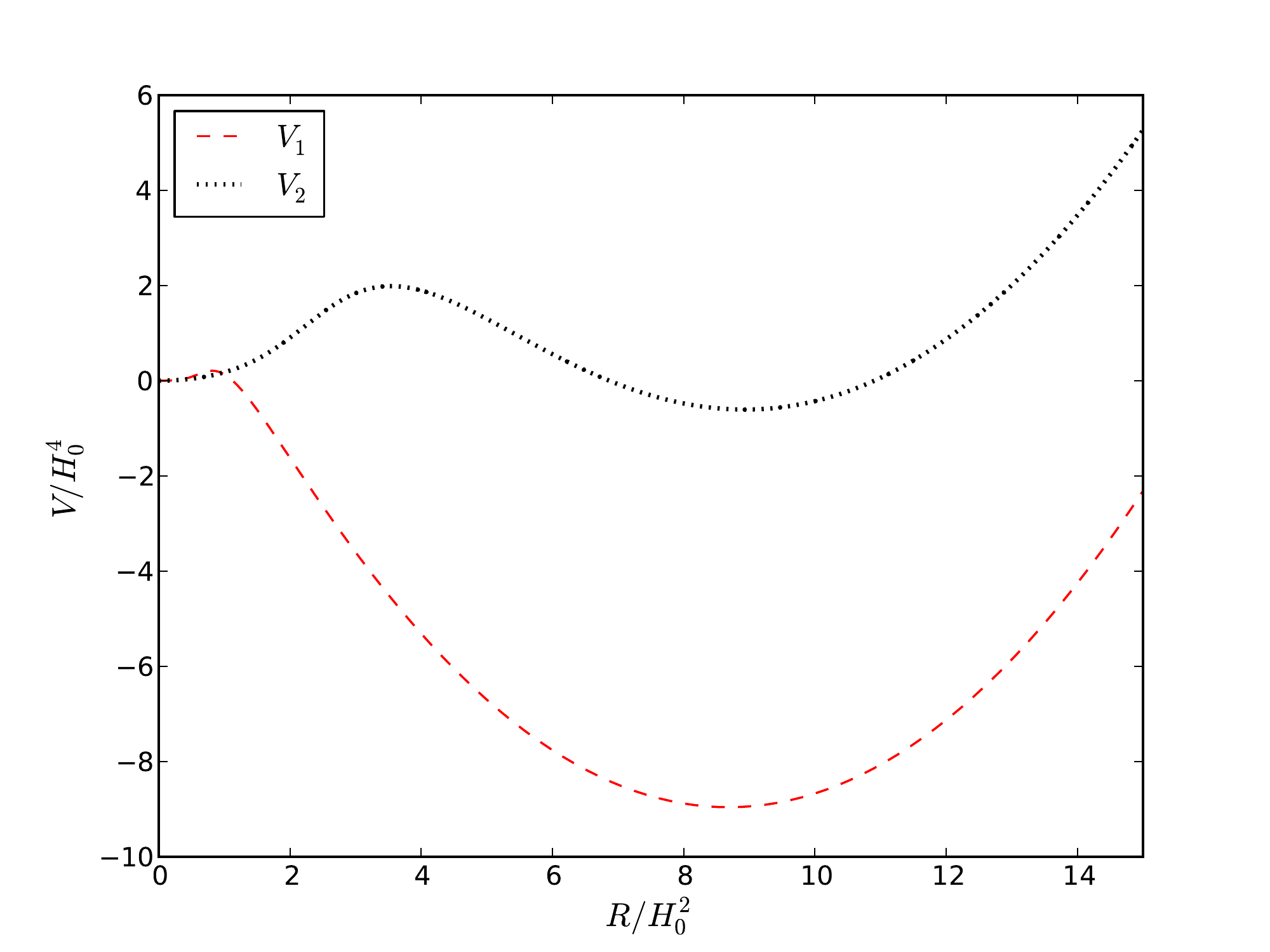}$
 \caption{\label{pot_hu} The potential $V(R)$ for the two different cases considered in the Hu--Sawicki model.}
\end{figure}

Similarly to the previous model, both sets of parameters result in present-day values of $H$ and $R$ that closely agree. Since the minima of the respective $V(R)$ potentials, shown in Fig.~\ref{pot_hu}, are located at almost equal $R$, it is expected that the background evolution of the models does not disagree much in the distant future.

In this model, we have $\Lambda_{\textrm{eff}}^{\infty} \approx 2.2 H_{0}^{2}$ and $\Lambda_{\textrm{eff}}^{{\rm de\, Sitter}} \approx 2.2 H_{0}^{2}$ for the first set, and $\Lambda_{\textrm{eff}}^{\infty} \approx 2.3H_{0}^{2}$ and $\Lambda_{\textrm{eff}}^{{\rm de\, Sitter}} \approx 2.2H_{0}^{2}$ for the second set of parameters. As for $\tilde{f}_{R0}$, we have $\tilde{f}_{R0} \approx -1\times 10^{-4}$ for the first set of parameters, and $\tilde{f}_{R0} \approx -1\times 10^{-2}$, recovering the result obtained in Ref.~\cite{frcosmo}. The stability of the solutions was guaranteed, as both $f_{RR}$ and $f_{R}$ remained definite positive throughout.

\begin{figure}[t!]
\begin{center}$
\begin{array}{c}
\includegraphics[scale = 0.40]{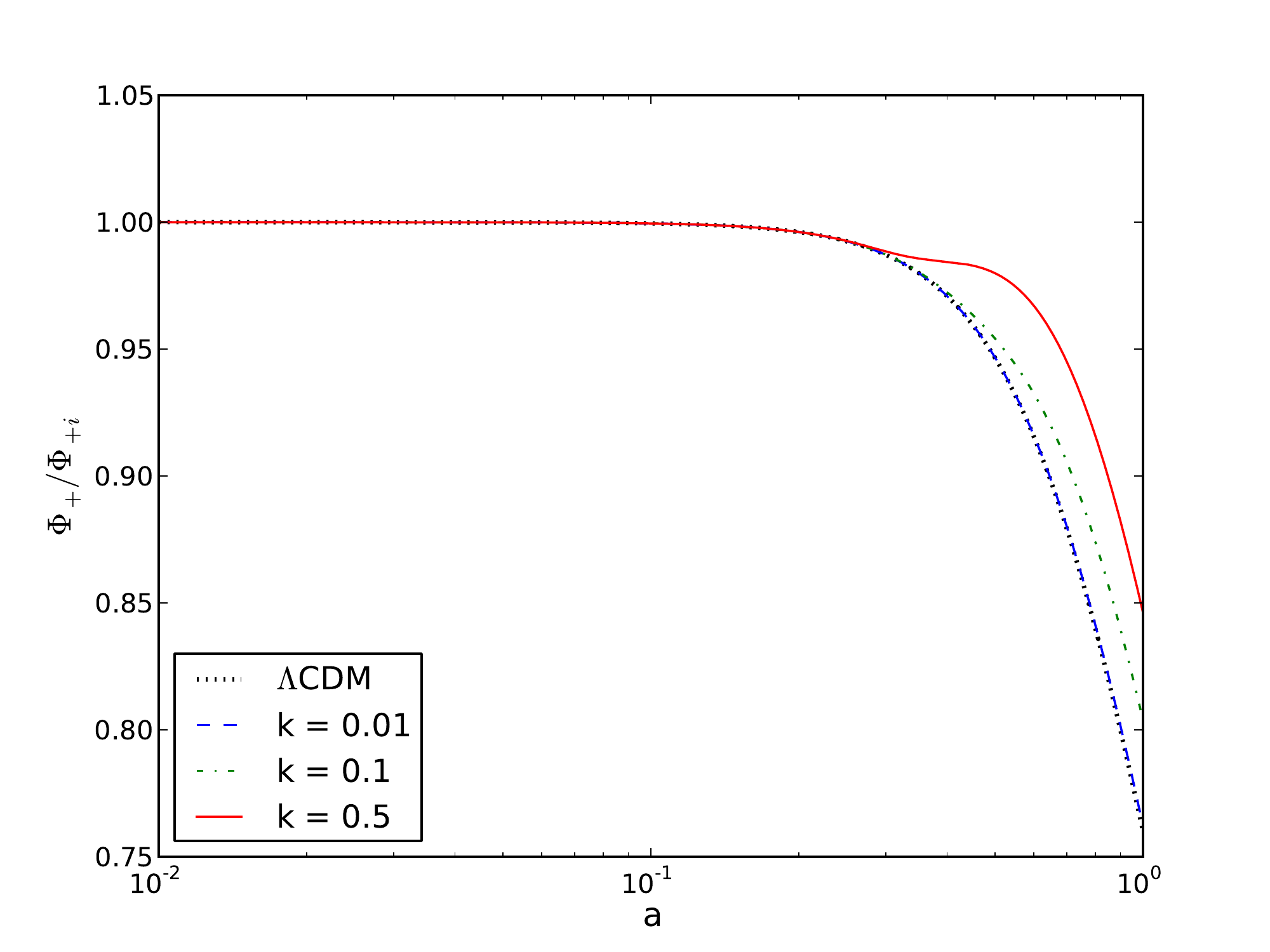} \\
\includegraphics[scale = 0.40]{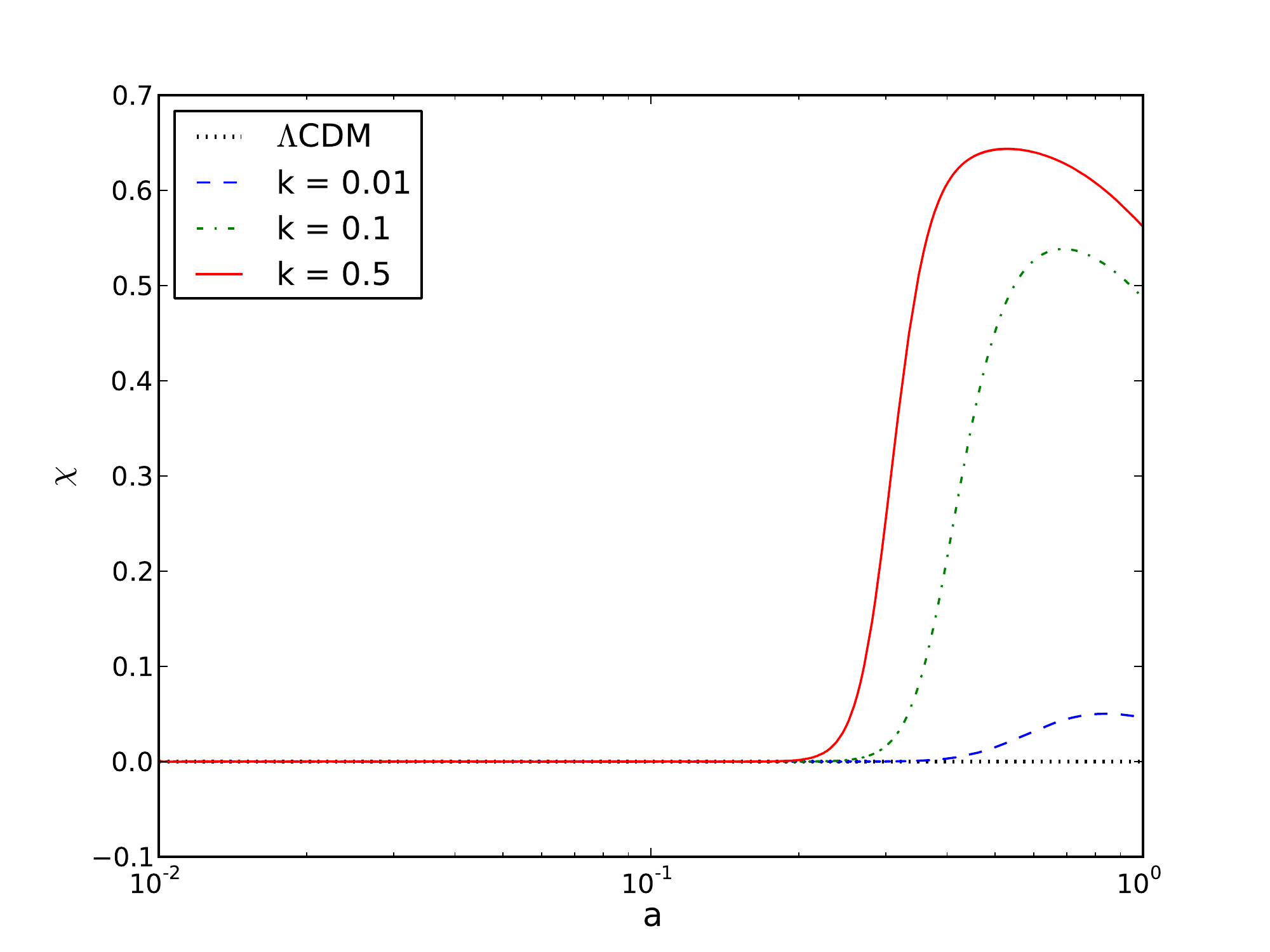}
\end{array}$
\end{center}
\caption{\label{perthu1}The lensing potential $\Phi_{+}$ and $\chi$ as a function of $a$ for the Hu--Sawicki model, for the first set of parameters.}
\end{figure}

\begin{figure}[t!]
\begin{center}$
\begin{array}{c}
\includegraphics[scale = 0.40]{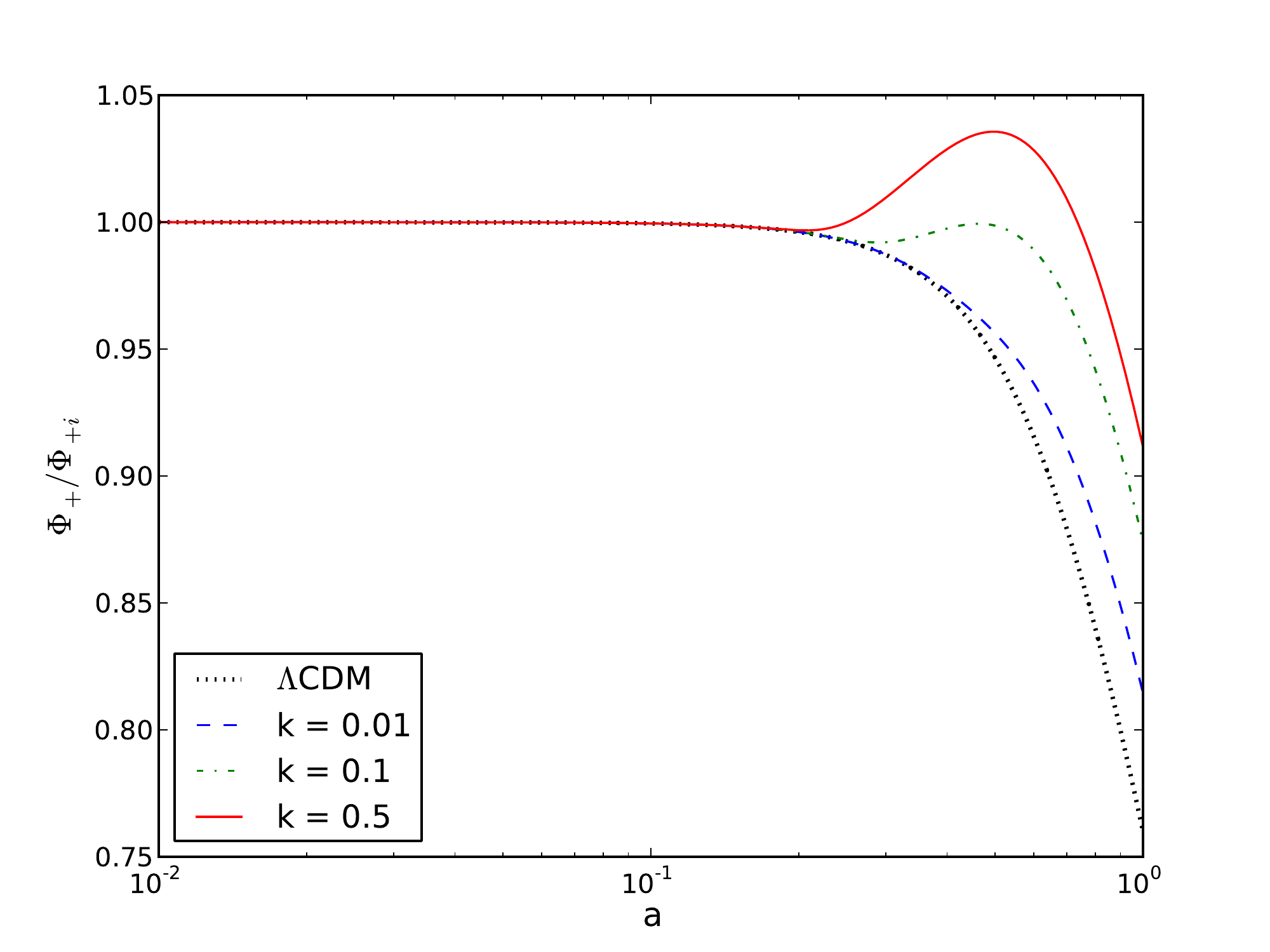} \\
\includegraphics[scale = 0.40]{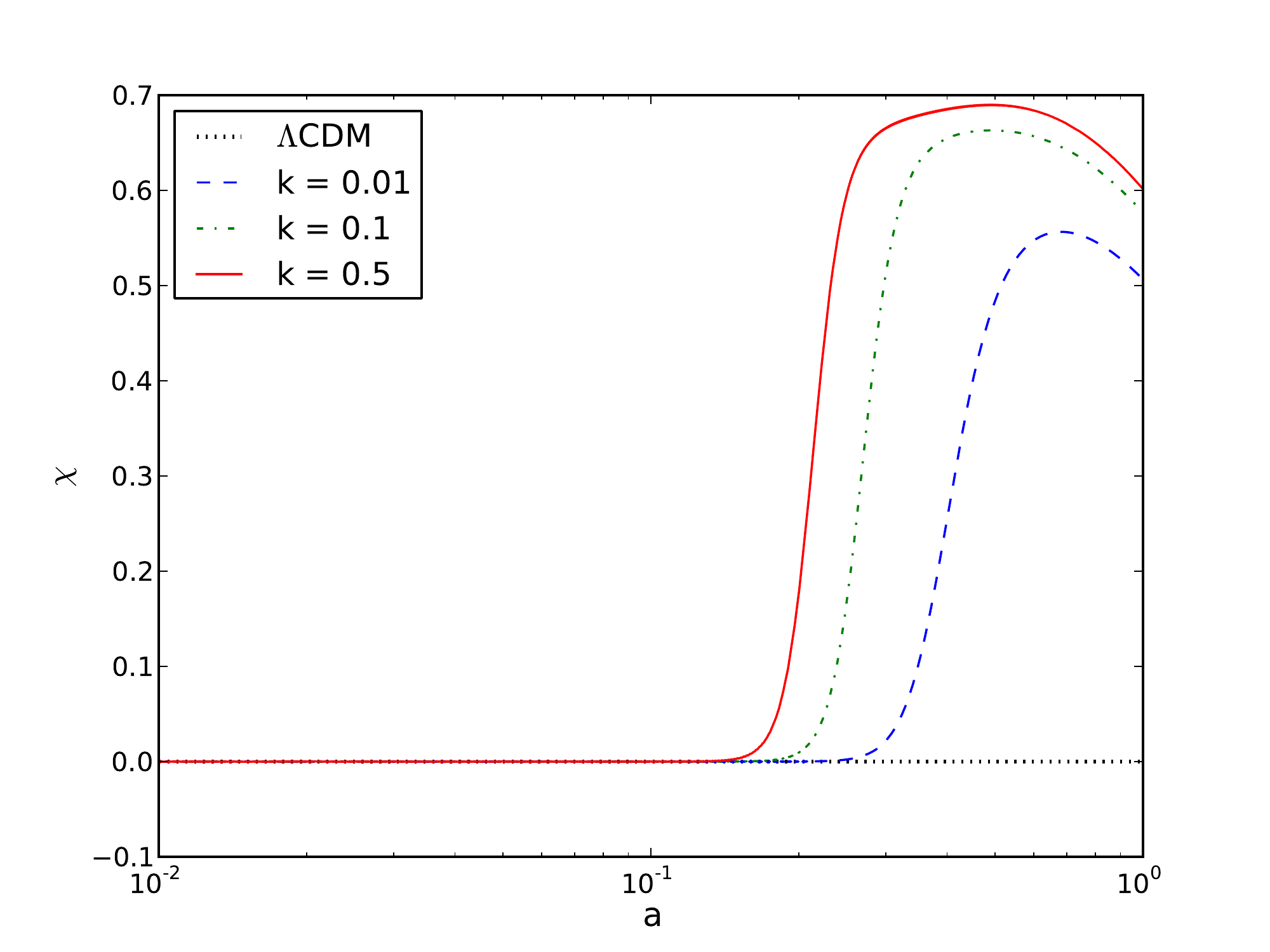}
\end{array}$
\end{center}
\caption{\label{perthu2}The lensing potential $\Phi_{+}$ and $\chi$ as a function of $a$ for the Hu--Sawicki model, for the second set of parameters.}
\end{figure}

\begin{figure}[t!]
 $\includegraphics[scale = 0.40]{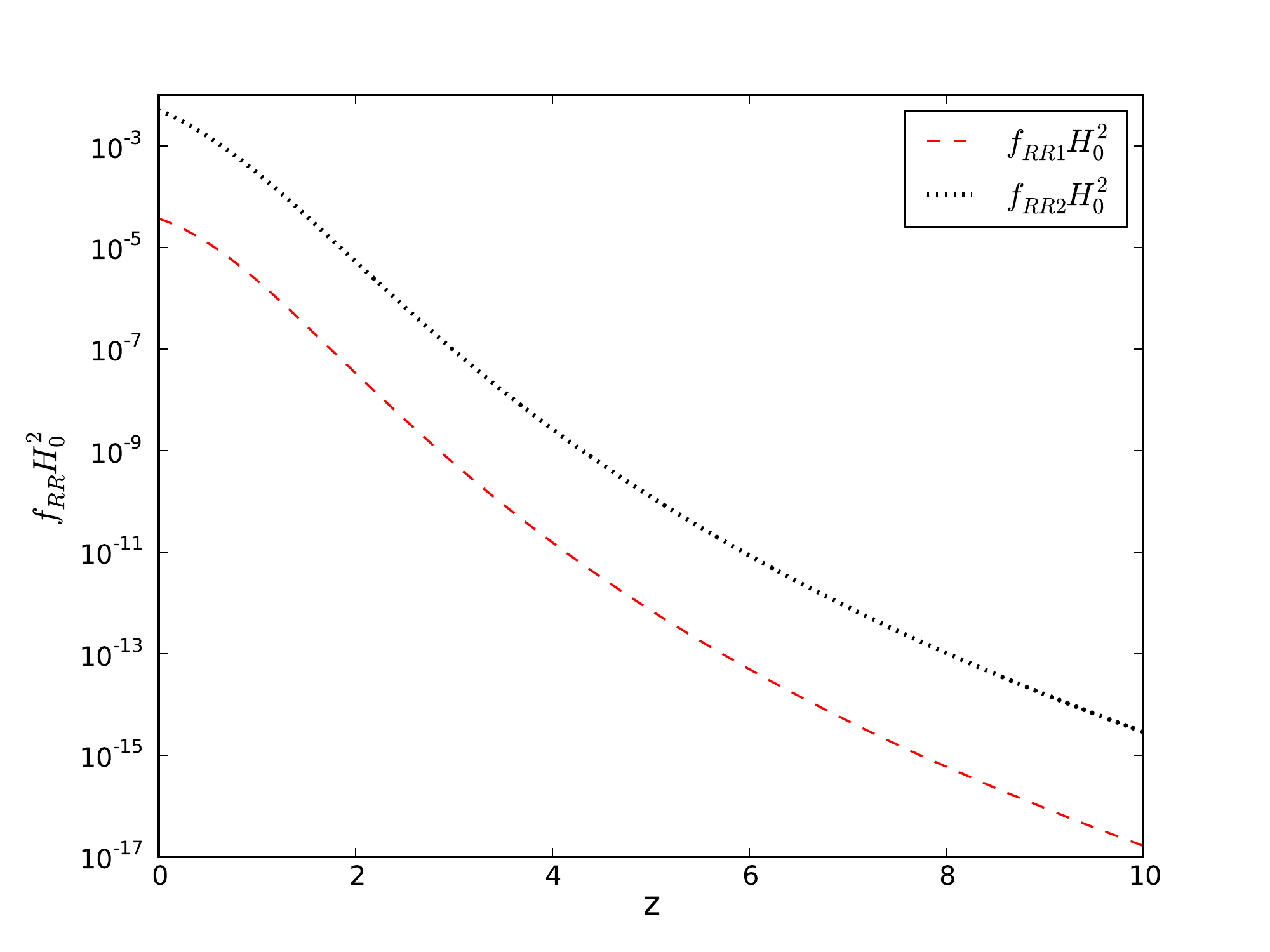}$
 \caption{\label{frr_hu} The form of $f_{RR}$ as a function of redshift for the two cases considered in the Hu--Sawicki model.}
\end{figure}

Figures~\ref{perthu1} and \ref{perthu2} show that the linear evolution of perturbations in this model is almost identical to the previous model, the reason being the similarity in the evolution of $f_{RR}$ of both models, seen for this model in Fig.~\ref{frr_hu}. The subtle difference rests on the absolute value of this quantity, which is smaller for the Hu--Sawicki model. Therefore, the different modes will enter the range of the fifth force marginally later. Hence, the enhancement of the gravitational potentials is a bit smaller than in the Starobinsky model, which is noticeable when comparing Figs.~\ref{perthu2} and \ref{pertstaro2}.

\subsection{\label{model3}Exponential model}

\begin{figure}[t!]
\begin{center}$
\begin{array}{c}
\includegraphics[scale = 0.40]{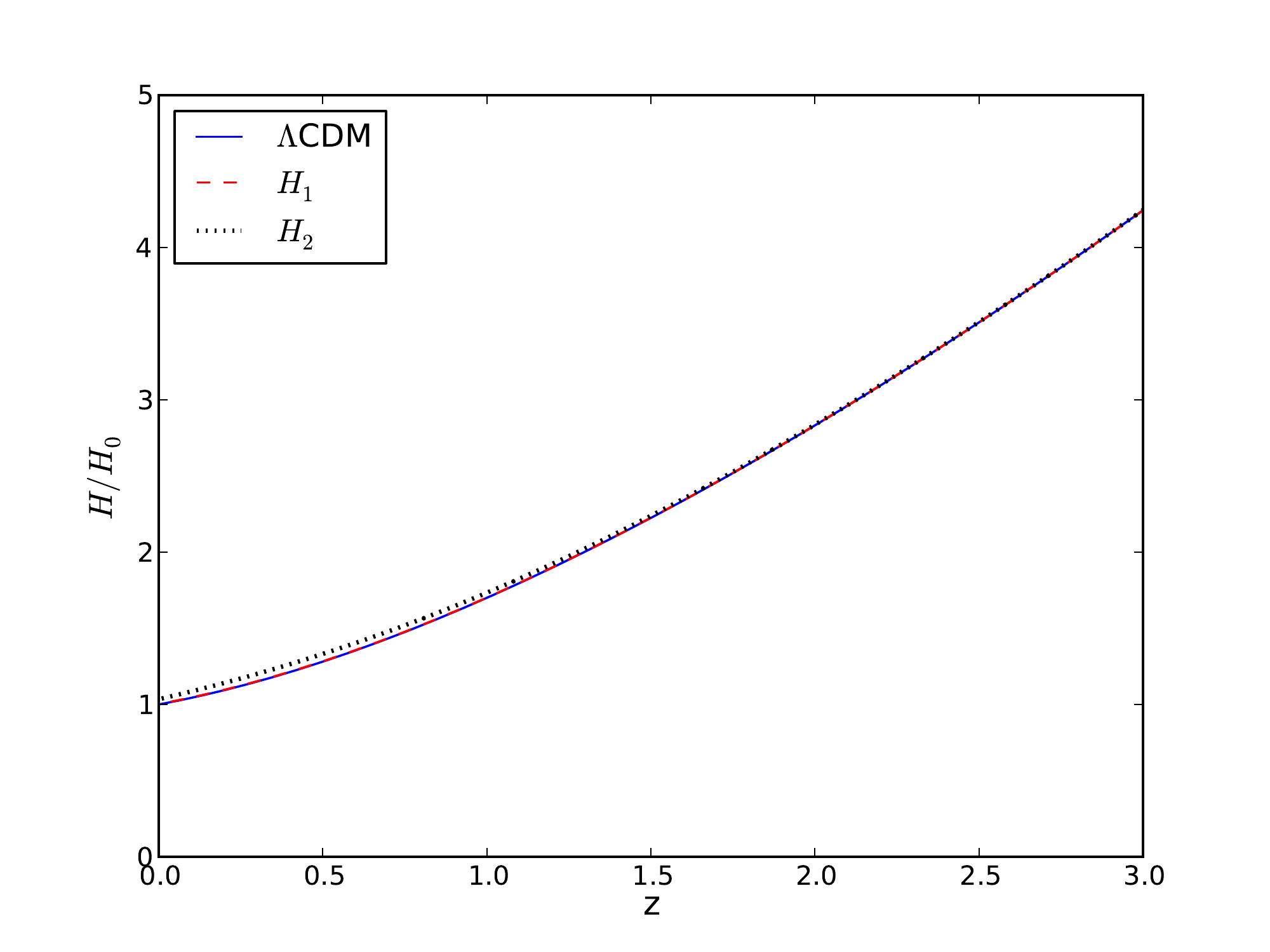} \\
\includegraphics[scale = 0.40]{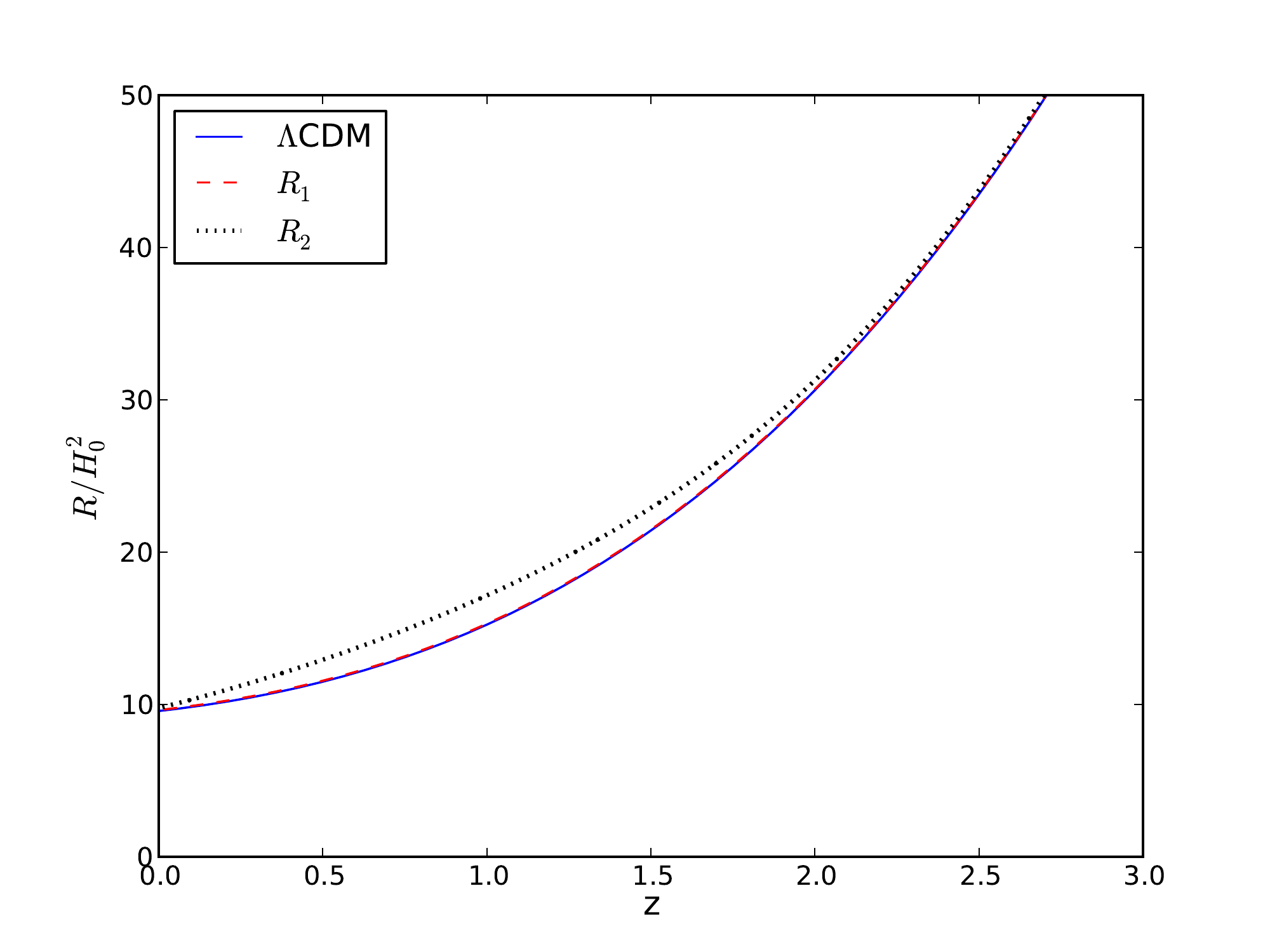} \\
\includegraphics[scale = 0.40]{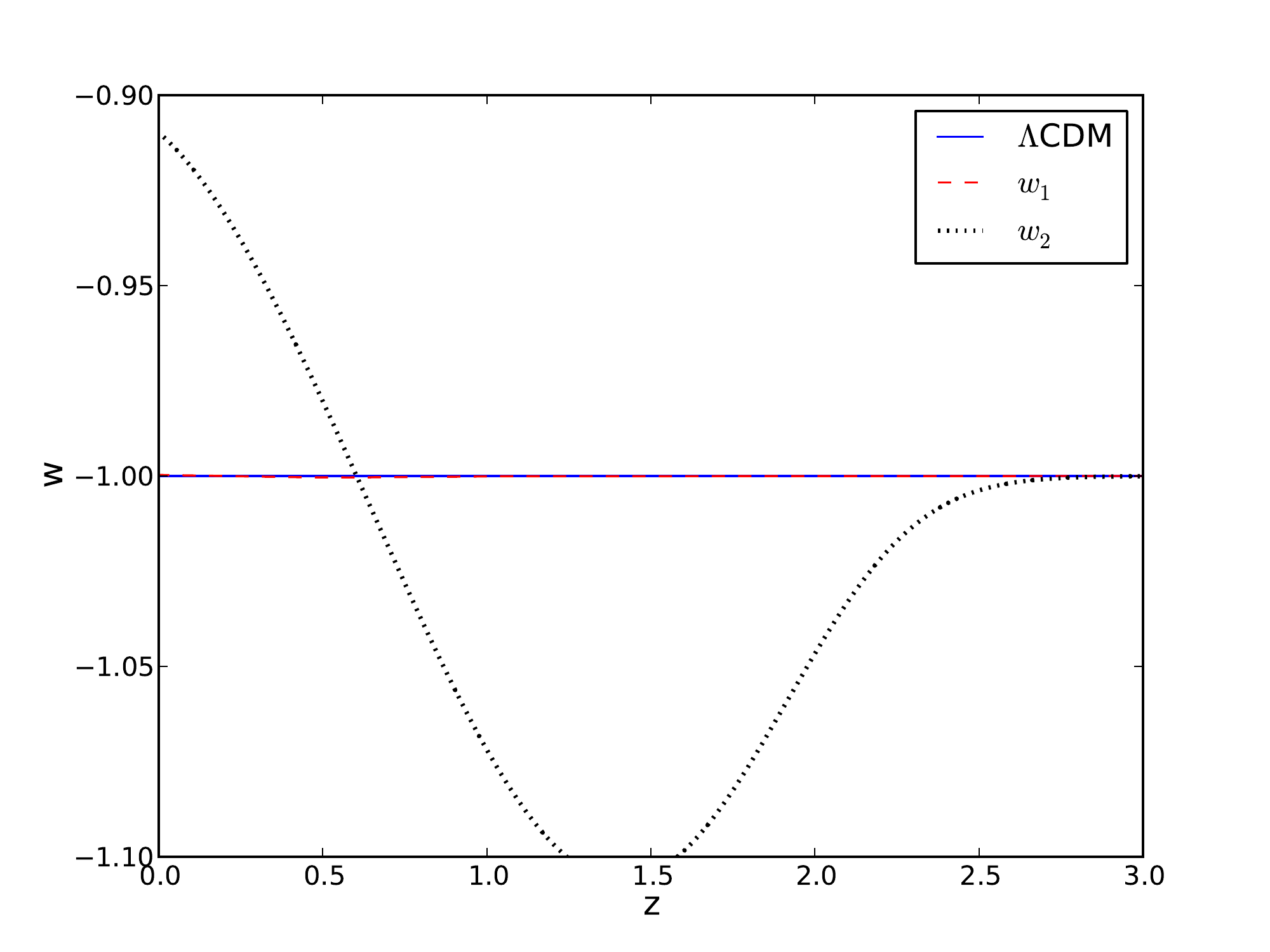}
\end{array}$
\end{center}
\caption{\label{backexpo}The Hubble parameter, Ricci scalar, and $f(R)$ effective equation of state for the Exponential model, compared to the $\Lambda$CDM model.}
\end{figure}

For the Exponential model, we have chosen $\lambda_{1} = 4.9$ and $R_{\star 1} = 0.9$, while $\lambda_{2} = 1.5$ and $R_{\star 2} = 3.0$. We plot the evolution, as a function of redshift, of the background quantities $H$, $R$ and $w_{\textrm{eff}}$, for both sets of parameters, in Fig.~\ref{backexpo}. As in the previous models, the cases considered yield background histories very close to $\Lambda$CDM, with the exception of the effective equation of state. The latter presents the phantom crossing mentioned before, once again more noticeable for the second set of parameters, and negligible for the first.

\begin{figure}[t!]
 $\includegraphics[scale = 0.40]{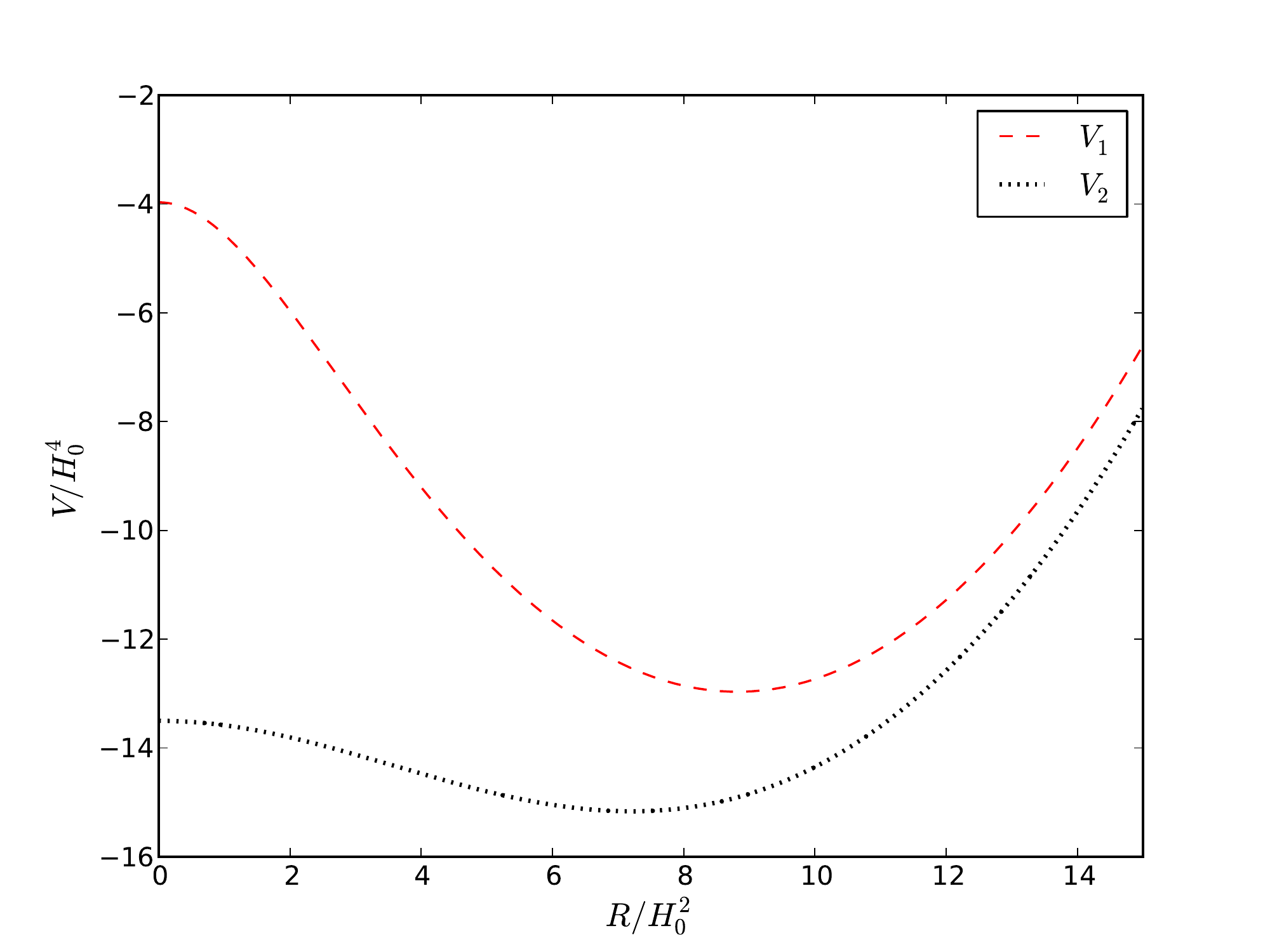}$
 \caption{\label{pot_expo}The potential $V(R)$ for the two cases considered in the Exponential model.}
\end{figure}

Contrasting with the previous models, the present-day values of $R$ and $H$ obtained for the first set of parameters of this model are very close to the respective values of $\Lambda$CDM and the second set of parameters. This happens because, when evolution ends, the solution for $R_{1}$ is already close to the respective $V(R)$ potential's de Sitter minimum, which is located at a higher value than that of the second set of parameters, as one can see in Fig.~\ref{pot_expo}. Since $R_{1}$ is expected to asymptotically settle at this value, one expects that, in the distant future, the respective evolution gradually differs from $R_{2}$, whose $V(R)$ minimum is located at a smaller value of $R$, and also differ from $\Lambda$CDM.

As for the values of the effective cosmological constants, for this model we have, considering the first set of parameters, $\Lambda_{\textrm{eff}}^{\infty} = 2.2 H_{0}^{2}$ and $\Lambda_{\textrm{eff}}^{\textrm{de\, Sitter}} \approx 2.2 H_{0}^{2}$, while for the second set we have $\Lambda_{\textrm{eff}}^{\infty} = 2.2 H_{0}^{2}$ and $\Lambda_{\textrm{eff}}^{\textrm{de\, Sitter}} \approx 1.8 H_{0}^{2}$, in accordance with Ref.~\cite{expo}. As for the values of $\tilde{f}_{R0}$, we have ensured that the first set of parameters results in a value of approximately $-1 \times 10^{-4}$, while the second set results in approximately $-6\times10^{-2}$.

\begin{figure}[t!]
\begin{center}$
\begin{array}{c}
\includegraphics[scale = 0.40]{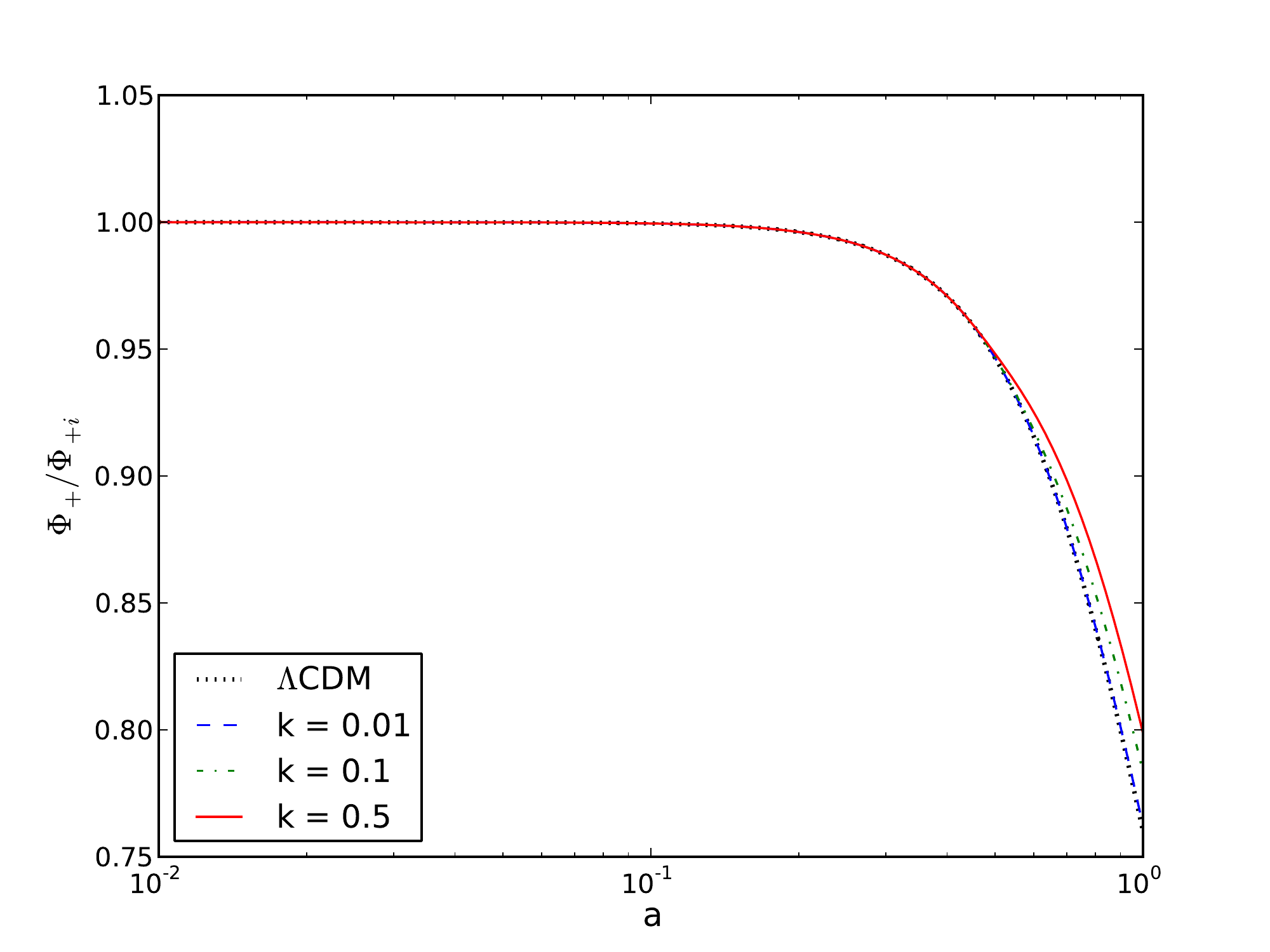} \\
\includegraphics[scale = 0.40]{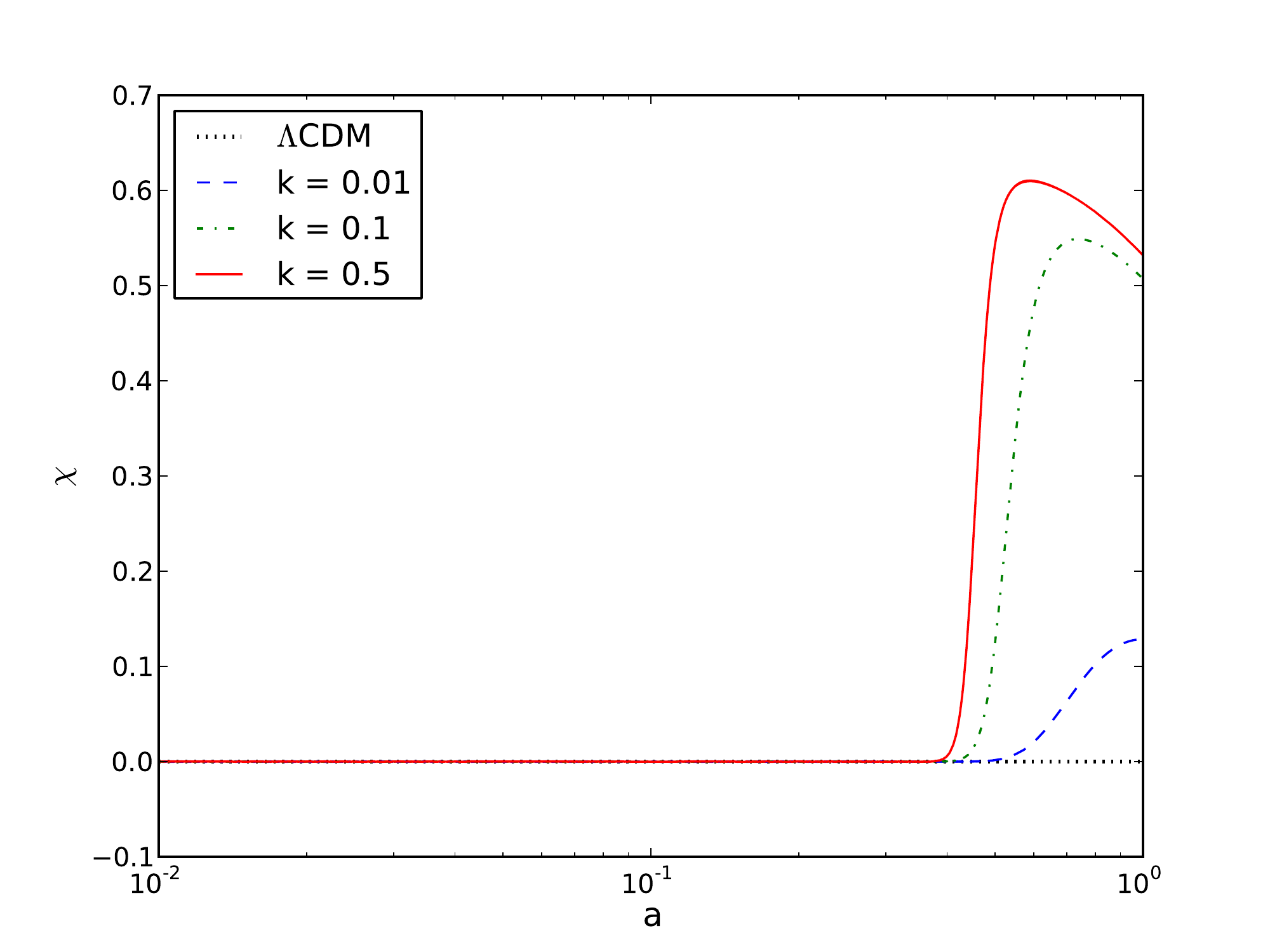}
\end{array}$
\end{center}
\caption{\label{pertexpo1}The lensing potential $\Phi_{+}$ and $\chi$ as a function of $a$ for the Exponential model, for the first set of parameters.}
\end{figure}

\begin{figure}[t!]
\begin{center}$
\begin{array}{c}
\includegraphics[scale = 0.40]{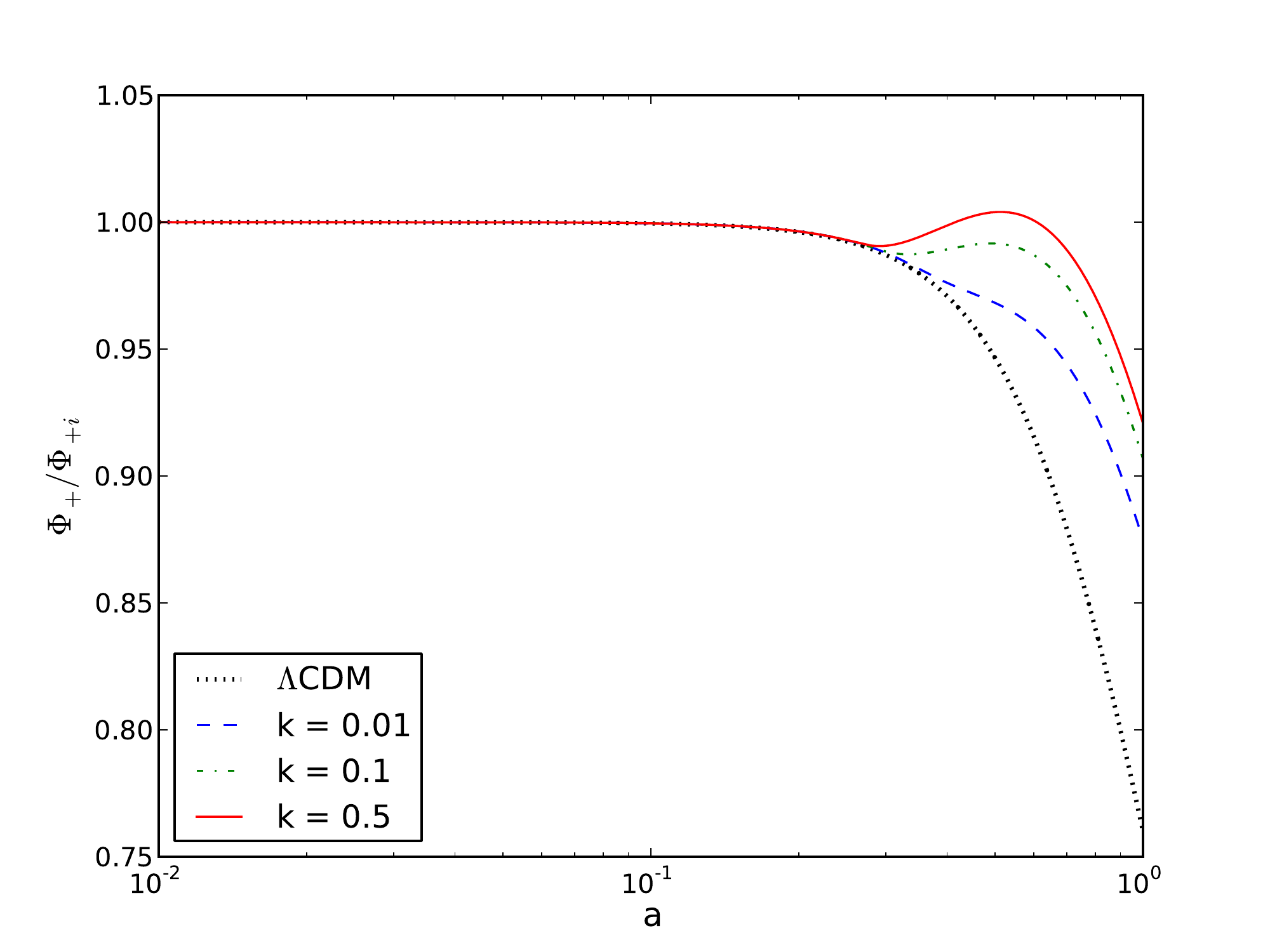} \\
\includegraphics[scale = 0.40]{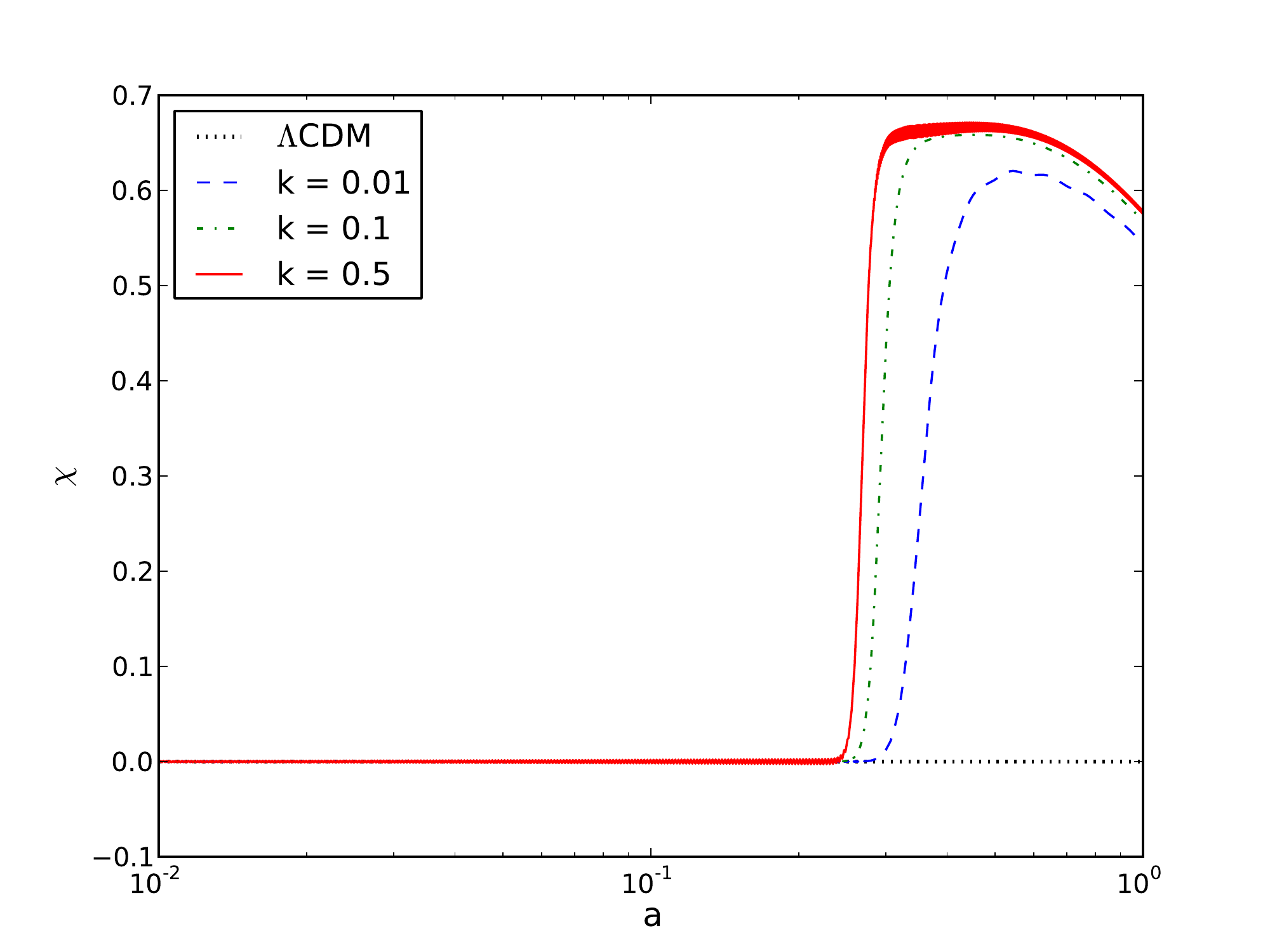}
\end{array}$
\end{center}
\caption{\label{pertexpo2}The lensing potential $\Phi_{+}$ and $\chi$ as a function of $a$ for the Exponential model, for the second set of parameters.}
\end{figure}

\begin{figure}[t!]
 $\includegraphics[scale = 0.40]{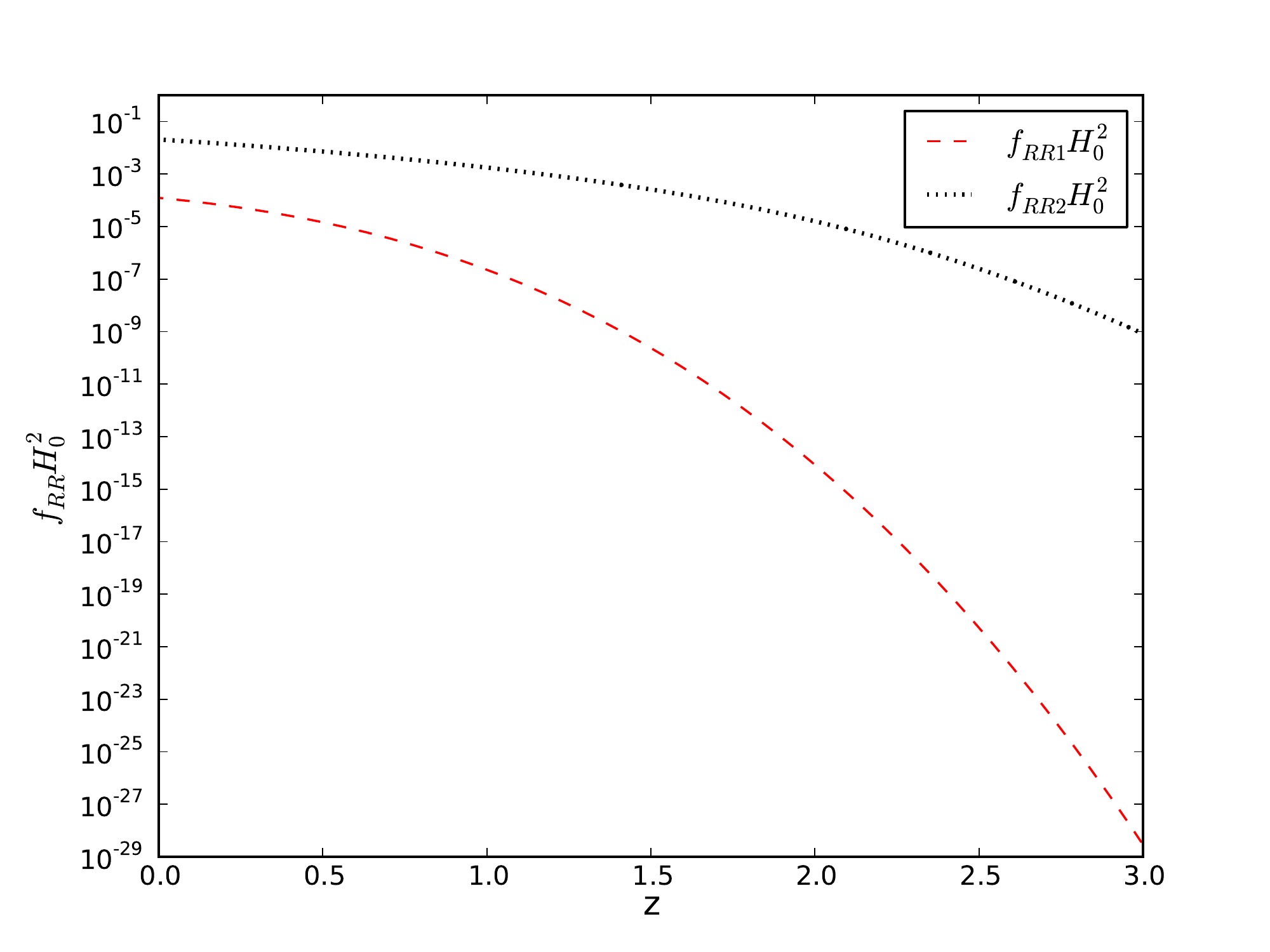}$
 \caption{\label{frr_expo}The form of $f_{RR}$ as a function of redshift for the two different cases considered in the Exponential model.}
\end{figure}

Regarding the evolution of linear perturbations, this follows a similar pattern to that of the previous models, as can be seen in Figs.~\ref{pertexpo1} and \ref{pertexpo2}. However, for the Exponential model, $f_{RR}$ is exponentially suppressed at high redshift. In Fig.~\ref{frr_expo} one sees that around $z=3$ the magnitude of $f_{RR}$ is still several orders of magnitude smaller than in the previous models. This means that the different $k$-modes will enter the fifth force range of action later and, consequently, the enhancement in the perturbations is much fainter for both sets of parameters. Note that, when we have $|\tilde{f}_{R0}| \approx 10^{-4}$, the difference between this model and $\Lambda$CDM is very small, even for the smallest scale considered. One final point regards the steep increase of $\chi$ particularly for the second set of parameters, which is related to the exponential growth of $f_{RR}$ for decreasing redshift. Therefore, even though the modes enter the range of the fifth force later,
 these do so at a very rapid pace, leading to an abrupt enhancement of the perturbations, namely $\chi$.

\subsection{\label{generic}$w = -1$ $f(R)$ model}

In this subsection, we present the evolution of the linearly-perturbed potentials for an $f(R)$ model with an effective equation of state identically equal to $-1$ throughout the cosmological evolution. Using the designer approach, we ensure that this model's background evolution is indistinguishable from the $\Lambda$CDM. We also tune the model such that the present-day value obtained for $\tilde{f}_{R0} \approx -1\times10^{-6}$. Hence, with this final model, we want to give an indication of what the evolution of the linear perturbations would be under more stringent viability constraints.

\begin{figure}[t!]
\begin{center}$
\begin{array}{c}
\includegraphics[scale = 0.40]{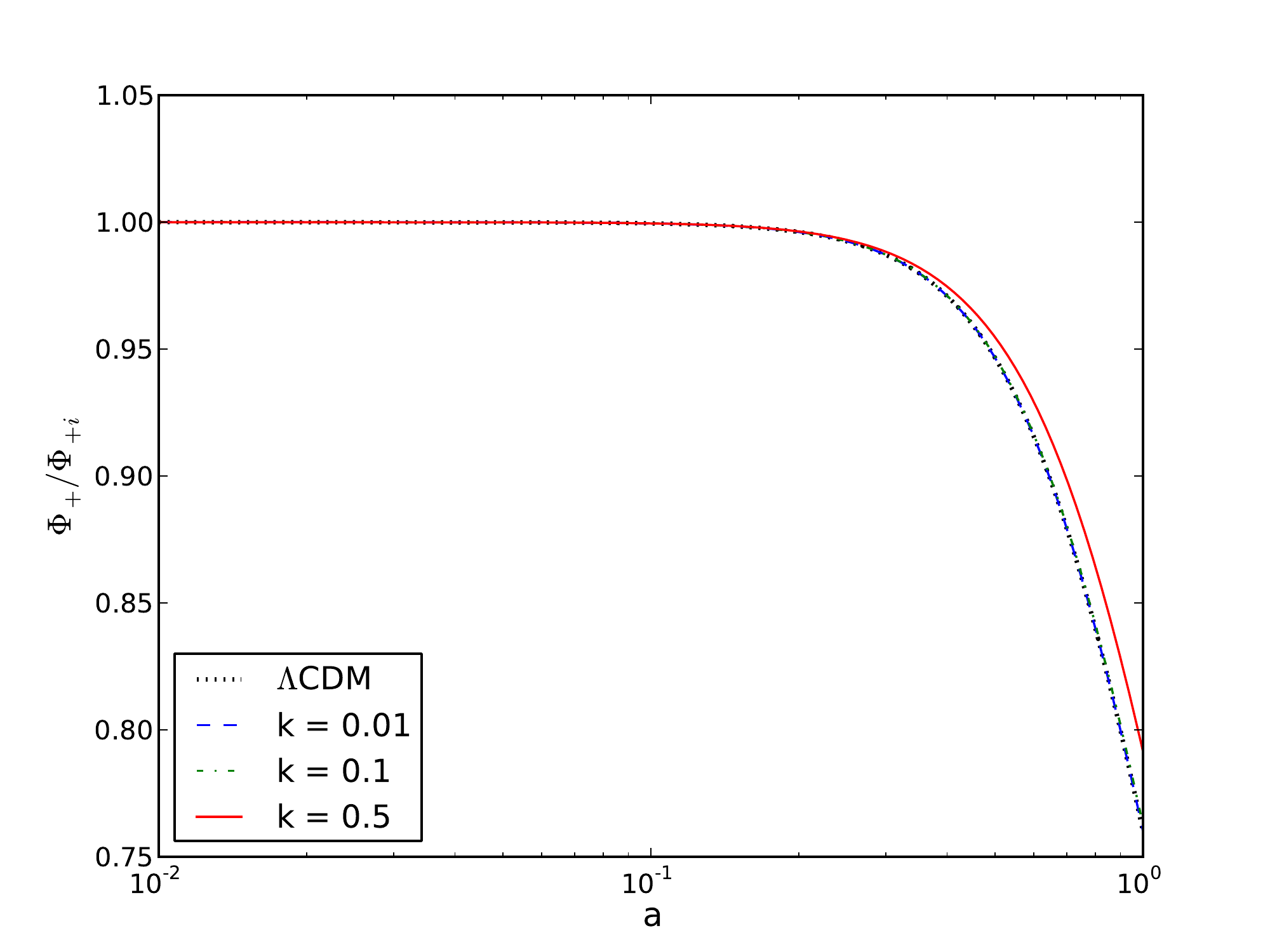} \\
\includegraphics[scale = 0.40]{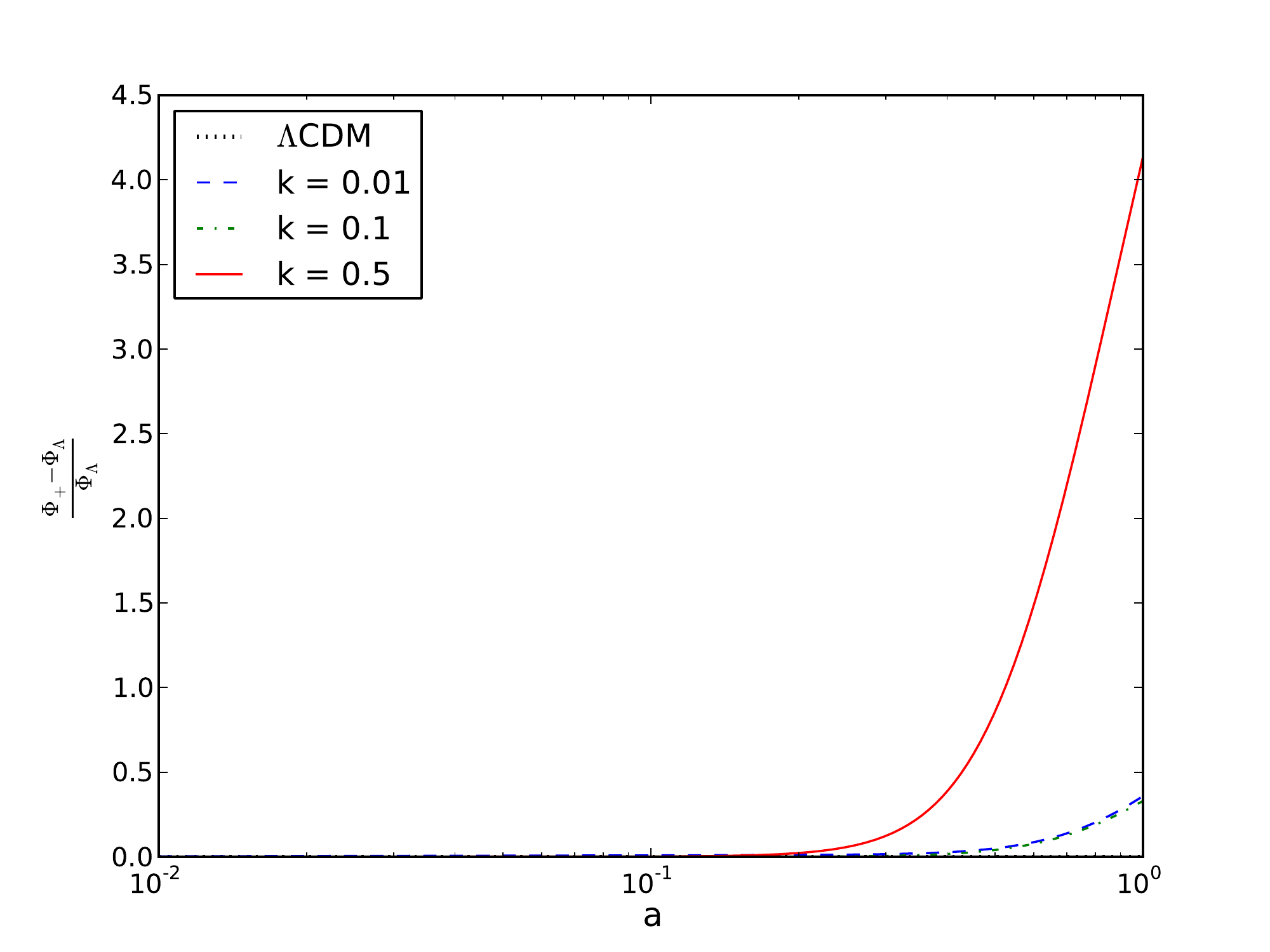} \\
\includegraphics[scale = 0.40]{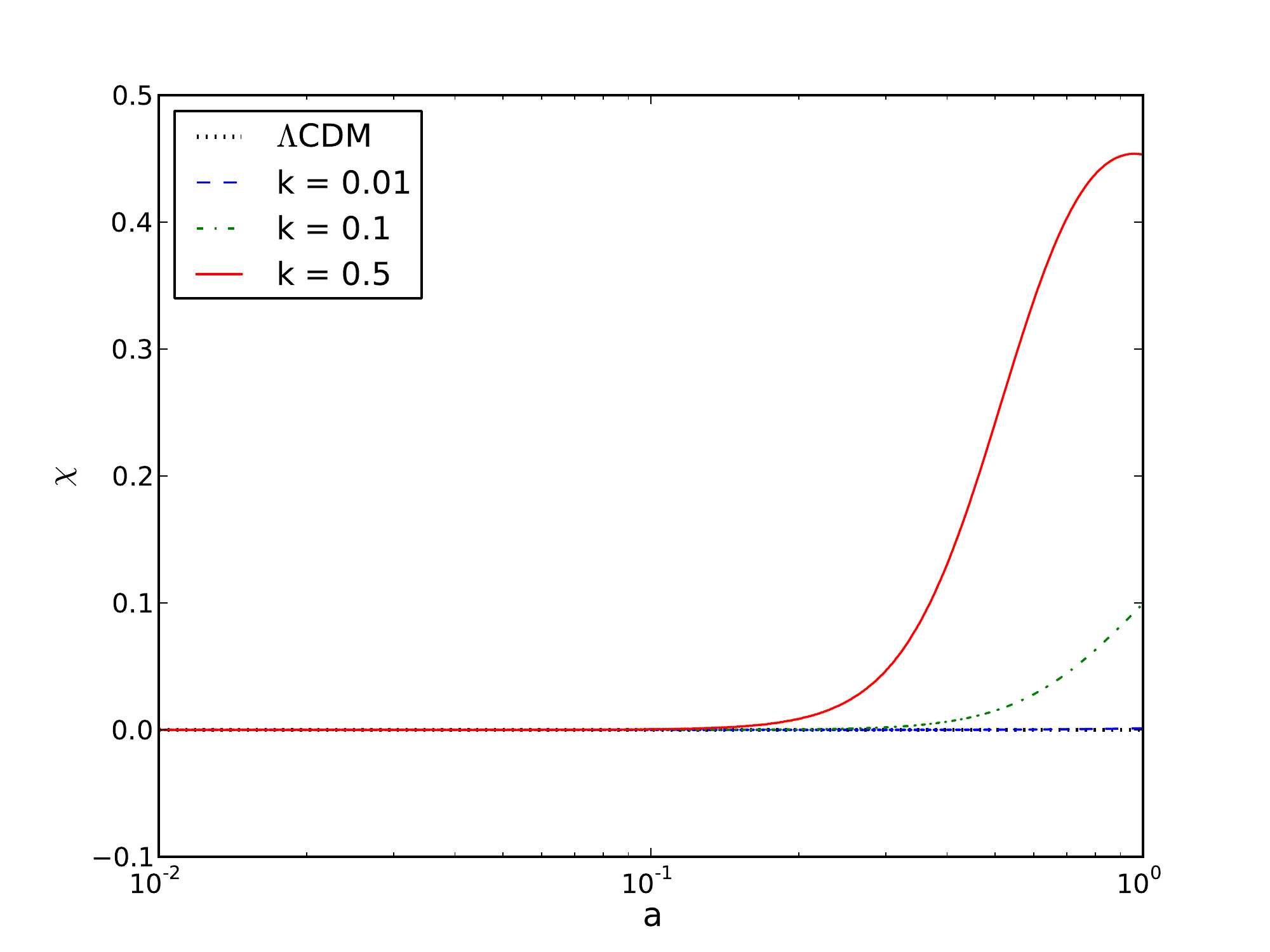}
\end{array}$
\end{center}
\caption{\label{pert_generic} The evolution of the linear perturbations as a function of the scale factor, $a$, for the $f(R)$ model with $w_{\textrm{eff}} = -1$, against the $\Lambda$CDM model. The middle plot shows the relative difference between both models in the evolution of $\Phi_{+}$. The result for $k = 0.01 h/{\rm Mpc}$ was enhanced by a factor of $100$ to allow its visualization.}
\end{figure}

Figure \ref{pert_generic} shows the evolution of the perturbations. One immediately observes that the enhancement in the lensing potential is almost negligible relative to the $\Lambda$CDM case. Only for the smallest scale can one detect by eye the difference between the models. Looking at the middle plot in Fig.~\ref{pert_generic}, one sees that the difference between the $f(R)$ model and $\Lambda$CDM in $\Phi_{+}$ ranges from $0$ to a maximum to $4\%$. So, even though $\chi$ does present some enhancement, albeit smaller than in all of the previous cases, that does not translate to the observable lensing potential $\Phi_{+}$.

\section{\label{conclusion}Conclusion}

In this work, we have focused on three viable $f(R)$ models. In spite of several, perhaps catastrophic, problem hovering over these theories regarding particle production and their weak-field limit \cite{staro2,egtreview,prob1,prob2}, they still receive a lot of attention because they provide insight into how simple modifications of the gravitational action can lead to departures from standard General Relativity.

We have used two different parameterizations for each of the three models considered. One of them had already been used in previous work. We have recovered the obtained results which, inevitably, render them nonviable, even though these present cosmological evolutions that are very close to $\Lambda$CDM. The values obtained for $|\tilde{f}_{R0}|$ are way above the viability requirement. If this were the case, the linear evolution of perturbations would have definite signatures that would probably have been already observed, since the departure from $\Lambda$CDM is very accentuated. In this situation, the action of the fifth force is significant, and one can indeed observe a great enhancement of the perturbed potentials at the late stages of evolution, with even some growth on the lensing potential $\Phi_{+}$, particularly on the smaller scales. Closer to the present, all of the scales end up succumbing to the effect of the expanding background and we see an inversion of the growth.

On the other hand, we have tried to fine-tune the other set of parameters such that the present-day value resulting from the cosmological evolution of the models would be within the observationally viability range, such that $\tilde{f}_{R0} \approx -10^{-4}$. In this case, we notice that the enhancement in the perturbations is more subtle for the Starobinsky and Hu--Sawicki model, and almost non-existent in $\Phi_{+}$ for the Exponential model. For the last model, even though the growth in $\chi$ remains, the lateness of this leads to almost no enhancement in $\Phi_{+}$, which offers close to no resistance to the background expansion. For the other models, one is able to detect some resistance to the background expansion in $\Phi_{+}$.

The main differences in the evolution of the gravitational potentials between the different models are dependent on the evolution of $f_{RR}$. The latter, in turn, is related to the effective mass of the scalar degree of freedom associated to $f(R)$, the scalaron, which defines the range of action of the fifth force. Hence, it determines the moment when the different scales enter its range and, therefore, are enhanced. However, despite the differences amongst the models, the possible observational signatures on the lensing potential become increasingly hard to detect compared to $\Lambda$CDM, particularly for the larger scales.

Lastly, we have considered the evolution of the perturbations for an $f(R)$ model with an effective equation of state $w_{\textrm{eff}} = -1$ and $\tilde{f}_{R0} \approx -10^{-6}$. This case is perfectly within both the observational and theoretical viability conditions and allows one to have an idea of the behavior of the gravitational potentials if all the three models respected these strict restrictions. For this case, we have used the designer approach to get a background history that is virtually indistinguishable from $\Lambda$CDM. 

In that last approach, we were able to conclude that the evolution of linear perturbations when the viability conditions are completely satisfied follows very closely that of $\Lambda$CDM. Even though there is the usual evolution and growth in $\chi$, this does not extend to $\Phi_{+}$. We note that for the largest scale considered, $k = 0.5 \, h/{\rm Mpc}$, one gets the largest deviations from $\Lambda$CDM, at a maximum only of approximately $4\%$. Note, however, that $f(R)$ simulations have shown that the linear approach does not work particularly well for these models on the smaller scales, specially in those cases where the magnitude of the fifth force is smaller, hence for smaller $|\tilde{f}_{R0}|$. Nevertheless, the high non-linearity of the $f(R)$ equations seems to further suppress the effect of the fifth force and the deviations from $\Lambda$CDM \cite{simulation}.

Therefore, if indeed it is a particular $f(R)$ model driving cosmic acceleration, it may be extremely hard to extract any signature of it, since the observational precision, for instance in weak lensing experiments, available today and in the near future, will not allow detection of such a signal from $\Lambda$CDM.

\acknowledgments

N.A.L.\ acknowledges financial support from Funda\c{c}\~{a}o para a Ci\^{e}ncia e a Tecnologia (FCT) through grant SFRH/BD/85164/2012. A.R.L.\ was supported by the Science and Technology
Facilities Council [grant number ST/K006606/1]. 
We would like to thank Alvaro de La Cruz-Dombriz, Scott Dodelson, Baojiu Li, and Bruno Moraes for useful comments on this paper.

\bibliography{biblio}

\end{document}